\documentclass[journal]{IEEEtran}
\ifCLASSINFOpdf

\else

\fi
\usepackage{amsmath}
\usepackage{makeidx}  
\usepackage{graphicx}
\usepackage{subfigure}
\usepackage{epstopdf}
\usepackage{bm}
\usepackage{cite}
\usepackage{stfloats}
\usepackage{multirow}   
\usepackage{xcolor}
\definecolor{mycolor_g}{RGB}{44,162,95}
\definecolor{mycolor_y}{RGB}{254,178,76}
\definecolor{mycolor_r}{RGB}{221,28,119}
\definecolor{mycolor_b}{RGB}{49,130,189}
\usepackage{float} 
\usepackage{booktabs} 
\usepackage{algorithm}
\usepackage{algorithmic}
\usepackage{diagbox}

\usepackage{mathrsfs}
\usepackage{amssymb}
\usepackage{graphicx}
\usepackage{url}

\usepackage{tabularx,booktabs}
\newcolumntype{C}{>{\centering\arraybackslash}X} 
\usepackage{lipsum}

\usepackage{makecell} 

\usepackage{graphicx}
\usepackage{color}

\begin{document}

\title{Over-the-Air ODE-Inspired Neural Network \\
for Dual Task-Oriented Semantic Communications
}

\author{\IEEEauthorblockN{Mengbing~Liu, \textit{Graduate Student Member, IEEE}, Jiancheng~An, \textit{Member, IEEE}, \\ Chongwen~Huang, \textit{Member, IEEE},  
and Chau Yuen, \textit{Fellow, IEEE} }
 \thanks{Mengbing Liu, Jiancheng An, and Chau Yuen are with the School of Electrical and Electronics Engineering, Nanyang Technological University (e-mails: \tt{mengbing001@e.ntu.edu.sg}, \tt{jiancheng.an@ntu.edu.sg},  \tt{chau.yuen@ntu.edu.sg}).}
\thanks{ Chongwen Huang is with the College of Information Science and Electronic Engineering, Zhejiang University, Hangzhou 310027, China, and with International Joint Innovation Center, Zhejiang University, Haining 314400, China, and also with Zhejiang-Singapore Innovation and AI Joint Research Lab and Zhejiang Provincial Key Laboratory of Info. Proc., Commun. \& Netw. (IPCAN), Hangzhou 310027, China (e-mail: \tt{chongwenhuang@zju.edu.cn}).}
}

\maketitle
\begin{abstract}

Analog machine-learning hardware platforms promise greater speed and energy efficiency than their digital counterparts. Specifically, over-the-air analog computation allows offloading computation to the wireless propagation through carefully constructed transmitted signals. In addition, reconfigurable intelligent surface (RIS) is emerging as a promising solution for next-generation wireless networks, offering the ability to tailor the communication environment. Leveraging the advantages of RIS, we design and implement the ordinary differential equation (ODE) neural network using over-the-air computation (AirComp) and demonstrate its effectiveness for dual tasks. We engineer the ambient wireless propagation environment through distributed RISs to create an architecture termed the over-the-air ordinary differential equation (Air-ODE) network. Unlike the conventional digital ODE-inspired neural network, the Air-ODE block utilizes the physics of wave reflection and the reconfigurable phase shifts of RISs to implement an ODE block in the analog domain, enhancing spectrum efficiency. Moreover, the advantages of Air-ODE are demonstrated in a deep learning-based semantic communication (DeepSC) system by extracting effective semantic information to reduce the data transmission load, while achieving the dual functions of image reconstruction and semantic tagging simultaneously at the receiver. Simulation results show that the analog Air-ODE network can achieve similar performance to the digital ODE-inspired network.
Specifically, for the image reconstruction and semantic tagging task, compared with the analog network without the Air-ODE block, the Air-ODE block can achieve around 2 times gain in both reconstruction quality and tagging accuracy.

\end{abstract}


\IEEEpeerreviewmaketitle

\section{INTRODUCTION}
The Internet of Things (IoT) is pivotal in 
enabling seamless connectivity and automated
data transmission for devices equipped with sensors and communication technologies. 
Given the ubiquitous deployment of devices, the integration of artificial intelligence (AI) algorithms at the edge of the network becomes increasingly vital. 
Such integration promises to enhance IoT applications by merging wireless communication systems with machine learning (ML) algorithms, paving the way for more intelligent and efficient operations \cite{madakam2015internet, zhu2020toward, huang2024challenges}. 

Under this context, the requirement is for an access point (AP) to successfully receive data from all IoT devices before any computational operation can be performed. Unfortunately, this process, especially when the number of IoT devices is extremely large, is not spectrum-efficient. As a remedy, over-the-air computation (AirComp) has emerged as a promising technology to achieve ultra-fast wireless data processing and aggregation in IoT networks\cite{chen2018over}. To be more specific, by enabling the concurrent data transmissions from all IoT devices over the same radio channel and exploiting the waveform superposition property of multiple-access channels (MACs) at the AP, the communication and computation processes can be achieved at the same time \cite{nazer2007computation,zhu2018mimo,liu2023over}. 
More recently, AirComp has begun to be adopted in federated learning (FL) systems for fast global model aggregation via exploring the superposition property of the wireless MACs, including optimizing device selection and beamforming for enhanced participation \cite{yang2020federated}, accelerating the convergence rate \cite{amiri2020machine,zhu2020one}, and improving both training accuracy and convergence speed\cite{cao2022transmission}.

But naturally, the computational performance of AirComp may be significantly affected by the propagation conditions between the transceivers due to physical obstacles. 
The past years have witnessed the integration of reconfigurable intelligent surfaces (RISs) into diverse communication systems, notably enhancing the quality of service (QoS), expanding network capacity, and simultaneously reducing energy consumption \cite{huang2019reconfigurable, di2020smart,wu2021intelligent,basar2024reconfigurable}. In response to the challenges in RIS-aided communication systems, substantial research has been conducted, including channel estimation \cite{wei2021channel,liu2022deep}, beamforming design \cite{huang2021multi,liu2023cooperative, TCOM_2022_An_Low}, and resource management \cite{yang2022federated}.
Diverging from these approaches that primarily optimize communication efficacy, some research has shifted focus toward phase optimization with computational objectives. Such work aims to improve the spatial degrees of freedom and enhance array gains in RIS-aided AirComp networks 
\cite{wang2020wireless,yu2020optimizing,fang2021over,zhang2022worst}.  In \cite{wang2020wireless}, the authors aimed the minimization of AirComp distortion through the joint design of beamformers at the AP, the phase configuration at the RIS, and the transmit powers, which quantifies the AirComp distortion for fast wireless data aggregation and efficient battery recharging. Then, \cite{yu2020optimizing} explored the mean squared error (MSE) reduction in AirComp via multiple RISs in cloud networks, optimizing phase configurations and the detector at the server.  
 Both \cite{fang2021over} and \cite{zhang2022worst} focused on optimizing AirComp in unfavorable propagation channel conditions, with the former using the Mirror-Prox method and the latter developing an efficient algorithm considering channel state information (CSI) uncertainties.

 With the rapid expansion of deep learning as a dominant ML technique, some researchers have begun to explore the possibility of replacing the digital neural network with the analog computation process via physical mechanisms, like optical \cite{lin2018all}, acoustic \cite{hughes2019wave}, and RF signals \cite{sanchez2022airnn}. Additionally, by using metasurfaces in these analog systems, the individual meta-atoms can take on the role of neurons. Their electromagnetic responses are reconfigurable and can be retrained for different computing tasks as needed \cite{JSAC_2023_An_Stacked, WC_2024_An_Stacked, liu2025onboard}.
Unloading computation into the wireless domain within deep learning systems offers significant advantages to reducing latency by enabling real-time processing directly in the transmission medium. Additionally,  analog processing minimizes the need for digital conversion and complex computations on receivers, thereby lowering energy consumption and reducing computational demands on infrastructure. 

 {  
Deep learning-based semantic communication (DeepSC) systems have excelled across varied communication settings by prioritizing the transmission of information's essence \cite{yao2024semantic, Arxiv_2024_Huang_Stacked,guo2024distributed, guo2025multiscale, wang2025deep, li2023towards, jiang2022wireless, huang2022toward, yang2023energy, Du2023Semantic, Jiang2024RIS, zhao2024A}. For example, \cite{jiang2022wireless} introduced a semantic video conferencing network that only transmits key pixels, incorporates advanced error detection, and utilizes channel feedback to boost transmission efficiency. \cite{huang2022toward} developed a reinforcement learning-based semantic coding method that enhances image quality at low bit rates using a novel bit allocation and a generative adversarial network (GAN)-based decoder. Similarly, \cite{yang2023energy} explored resource allocation and semantic extraction for efficient communication, introducing an algorithm that optimizes computational and transmission energy, validated by numerical results. Recent research integrates RIS into SC systems to enhance communication efficiency and adaptability. In \cite{Du2023Semantic}, RIS is used to encode multiple signal spectrums into a single MetaSpectrum, reducing data volume and improving secure transmission. In \cite{Jiang2024RIS}, the RIS-SC framework dynamically allocates semantic content to adapt to user needs, preserving essential semantics while reconstructing lost non-essential details for clarity in diverse scenarios. Additionally, \cite{zhao2024A} shows how RIS enhances probabilistic semantic communication in industrial Internet-of-Things (IIoT) environments by dynamically optimizing phase shifts and power allocation to improve spectral efficiency and transmission rates, while intelligently managing semantic compression and user associations.
 }

While semantic communication represents a leap forward in reducing the volume of data needing transmission, it does not fully address the spectrum efficiency challenges in the traditional `transmit-then-compute' model. To bridge these two advancements, this paper proposes the innovative use of distributed RISs in semantic communication systems. { Moving beyond their traditional roles in enhancing link quality and spectrum efficiency \cite{Du2023Semantic, Jiang2024RIS, zhao2024A}, in this study, distributed RISs are strategically repurposed to act as functional elements of an analog neural network. This innovative application aims to perform complex computational tasks directly within the network infrastructure, such as data processing and pattern recognition, thus achieving computational goals that extend well beyond conventional communication objectives. This integration not only complements the semantic reduction of data transmission but also decreases latency and enhances computational efficiency by enabling a `compute-while-transmitting' capability in distributed RIS-aided communication systems. } 
Thus, the main contributions of this paper are concluded as 
\begin{itemize}
    \item   A complex-valued over-the-air ordinary
differential equation (Air-ODE) neural network is proposed for dual task-oriented semantic communications, mapping digital ODE-inspired neural networks to analog Air-ODE networks supported by distributed RISs to represent the coupled amplitude and phase information together in the RF domain.
    \item A DeepSC system is designed to extract effective semantic information to reduce the data transmission load while achieving the dual functions of image reconstruction and semantic tagging simultaneously at the receiver. Enhanced by distributed RISs, this setup enables computational offloading, resulting in more efficient transmission and processing for both tasks.
    \item  For practical deployment, the updates of the ODE block are implemented on distributed RISs taking into account discrete finite phase configurations, with the straight-through estimator (STE) employed to address gradient discontinuities caused by quantization during the training process of the digital ODE-inspired network.
    \item A two-stage fine-tuning algorithm is developed to balance disparate loss functions in image reconstruction and semantic tagging tasks. By initially focusing on image reconstruction and subsequently fine-tuning only the decoder for both tasks, this method ensures optimized results with satisfactory performance in both domains while reducing training costs.
    \item The performance of the analog Air-ODE neural network is assessed using peak signal-to-noise ratio (PSNR), structural similarity index (SSIM), and tagging accuracy. Simulation results indicate that performance improves with higher signal-to-noise ratios (SNR) and gradually approaches that of the digital ODE-inspired network. Compared to configurations without the Air-ODE block, the analog Air-ODE neural network achieves approximately twice the gain in all performance metrics.
    
\end{itemize}

\textit{Notation}:  
Uppercase (lowercase) boldface letters represent matrices (vectors). The operations of transpose and Hermitian transpose are indicated by $(\cdot)^T$ and $(\cdot)^H$, respectively.
The field of complex numbers is denoted by $\mathbb{C}$, and real numbers by $\mathbb{R}$. For a complex number, $\Re { (\cdot) }$ and $\Im { (\cdot) }$ represent the real and imaginary parts, respectively. The symbol $\mathbf{A} * \mathbf{B}$ denotes the convolution of matrices $\mathbf{A}$ and $\mathbf{B}$.
The notation ${\rm{diag}}(\mathbf{x})$ refers to a diagonal matrix formed with the elements of the vector $\mathbf{x}$. 
The element at the $m$-th row and $n$-th column of a matrix is represented as $\left[ \cdot \right]_{m,n}$. 
Additionally, $v \sim \mathcal{CN}(0,\sigma^2)$ indicates that $v$ follows a complex Gaussian distribution with zero mean and variance $\sigma^2$.

\section{System Model and Problem Formulation}
\begin{figure*}[hbt]
\centering
\includegraphics[scale=.38]{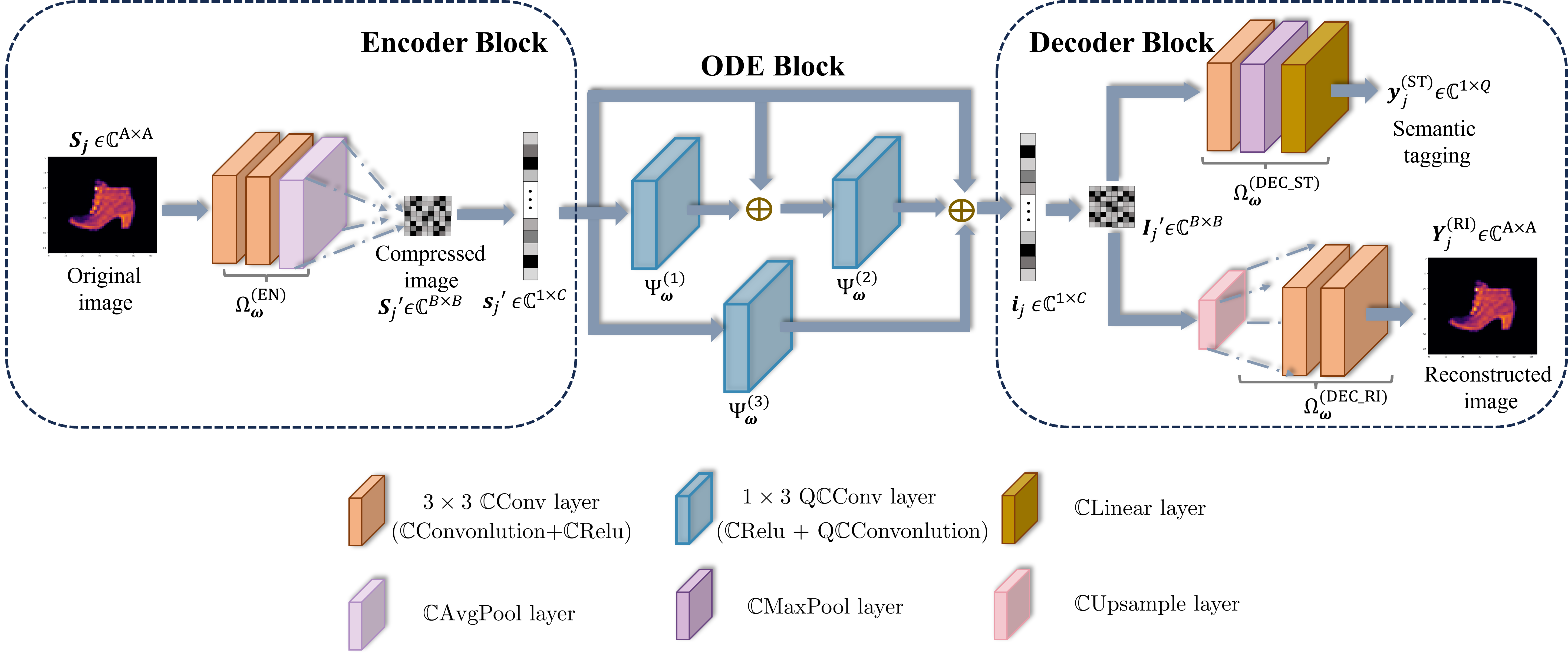}
\caption{The proposed complex-valued ODE-inspired neural network structure for the semantic communication system.}
\label{digital}
\end{figure*}
{ 
In this section, we introduce an analog Air-ODE as a replacement for the digital ODE-inspired neural network in semantic communication systems, aimed at enhancing image reconstruction and semantic tagging.  To mitigate the high computational complexity associated with digital operations, the ODE block is offloaded to wireless propagation, and this operation is efficiently executed through distributed RISs.   In subsequent sections, we will delineate the tasks, elucidate the relationship between the ODE block and the Runge-Kutta family, and detail the implementation of the Air-ODE block enabled by distributed RISs.
}

\subsection{Task Description}

{  Images possess both a rich visual and significant semantic space. Humans naturally process this information through brain activity \cite{gaziv2022self}, but devices struggle to interpret it directly, especially with compressed data.
To enhance efficiency, integrating these tasks into a single model is more effective than using separate models for training and inference. As shown in Fig. \ref{digital}, the encoder block compresses the image and extracts key information, which the ODE block further refines. The decoder then reconstructs the image, directing it to task-specific decoders with two distinct branches:
}
 \begin{itemize}
     \item For image reconstruction, the decoder block decodes the $j$-th image from its compressed form $\boldsymbol{I}_j'$. It learns to map from $\boldsymbol{I}_j'$ to the original $\boldsymbol{S}_j$, aiming to recover high-quality images that can generalize to unseen test data.
     \item  Additionally, beyond reconstructing images, the decoder also deciphers their semantic categories. It processes $\boldsymbol{I}_j'$ to extract semantic features, associating them with predefined categories $\boldsymbol{l}_j$.
 \end{itemize}
 
  By doing so, the model not only reconstructs the visual appearance but also interprets the semantic context, enhancing its applicability in various practical scenarios.
 Both the design of the encoder and decoder blocks need to be deployed at the transmitters and receivers, detailed processing functions will be given after introducing the digital ODE-inspired neural network in the next section.

\subsection{The Conventional ODE Structure}

This network architecture uses a digital ODE-inspired residual framework with direct paths for information flow between layers, enhancing the model's expressiveness and ability to capture intricate features, which is especially beneficial for handling image and semantic information \cite{he2016deep}.

{ 
 The theoretical foundations of ODEs emphasize the use of high-order approximations to improve accuracy metrics \cite{he2019ode}. Prominently, the Runge-Kutta methods, widely employed in numerical ODE solutions, are formulated as follows:
 \begin{equation}
    u_{n+1} = u_n + \sum_{i=1}^I \gamma_i \kappa f(v_n + \alpha_i \kappa,u_n +  \sum_{j=1}^{i-1} \mu_{ij} G_j), 
    \label{RK}
 \end{equation} 
 where \(v_n\) is the input at step \(n\), \(u_n\) is the system's state at that step, and \(u_{n+1}\) is the next state. Here, $I$ represents the number of stages; $\alpha_i$, $\mu_{ij}$ and $\gamma_i$ are the related coefficients of the $i$-th stage, $i = 1, 2, \cdots, I$.  For a two-stage process $I = 2$, setting  $\gamma_1 = \gamma_2 = 1/2$, $\alpha_1= 0$,  $\alpha_2= 1$ and $\mu_{21} = 1 $,  results in the Runge-Kutta equation we adopted:
\begin{align}
u_{n+1} &= u_n + \frac{1}{2} (G_1 + G_2), 
\end{align}
 where  $G_1$ and $G_2$ can be separately written as 
 \begin{align}
     G_1 &= \kappa f(v_n, u_n), \\ 
     G_2  &= \kappa f(v_n + \kappa, u_n + G_1).
     \label{G}
 \end{align}
{  Here, $G_1$ represents the initial increment calculated from the system's state at the beginning of the interval, while $G_2$ provides an updated increment from a subsequent point, reflecting progressive adjustments in this 2-stage Runge-Kutta method.}
Similarly, to integrate this into an ODE block, we use the module { $\Psi_{\boldsymbol{\omega}}$}, constructed with complex-valued ReLU ($\mathbb{C}$ReLU) and complex-valued convolution ($\mathbb{C}$Conv) functions. Unlike traditional residual network designs, the ODE block features multiple branches, making it popular in recent network designs \cite{chen2018neural,he2019ode,xu2021ordinary}. With Eq. (\ref{RK})$\sim$(\ref{G}), the ODE block is constructed as shown in Fig. \ref{ODE}.
  \begin{figure}[ht]
\setlength{\abovecaptionskip}{0pt}
\setlength{\belowcaptionskip}{0pt}
\centering 
\includegraphics[width= 0.45\textwidth]{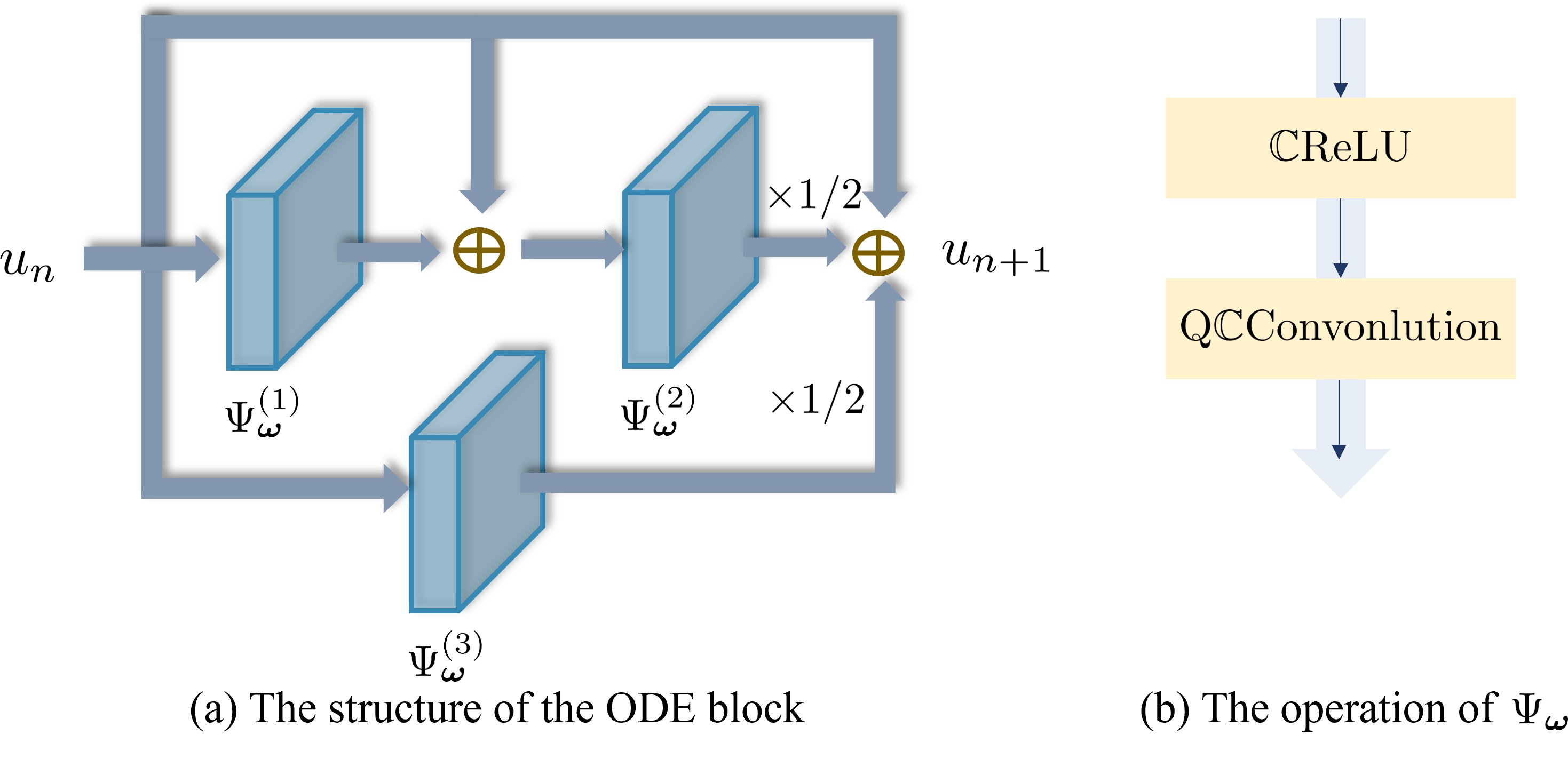}
\DeclareGraphicsExtensions.
\caption{The architecture of the ODE block.}
\label{ODE}
\vspace{-5pt}
\end{figure}
   \begin{figure}[ht]
\setlength{\abovecaptionskip}{0pt}
\setlength{\belowcaptionskip}{0pt}
\centering 
\includegraphics[width= 0.45\textwidth]{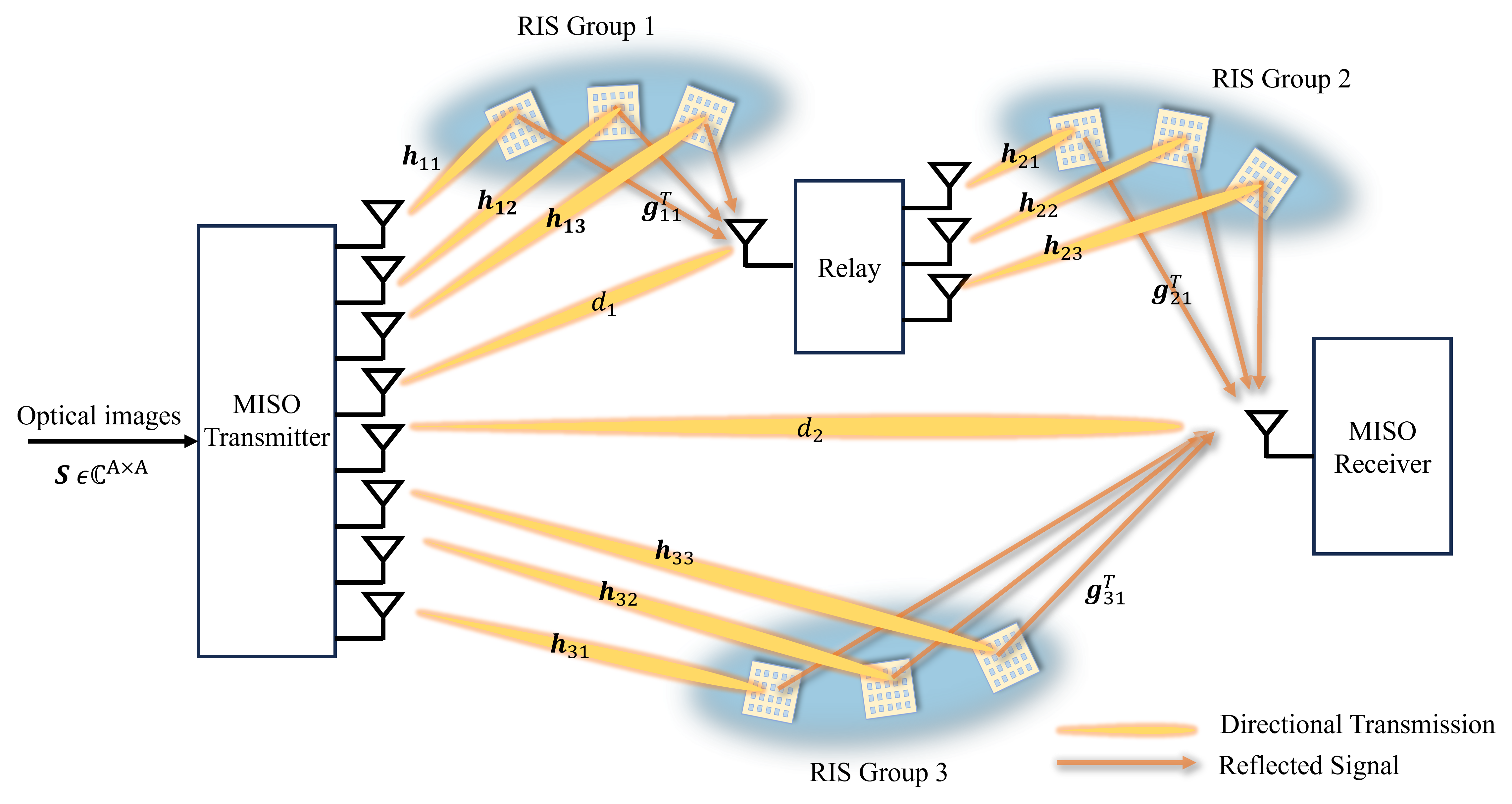}
\DeclareGraphicsExtensions.
\caption{Illustration of the distributed RISs-aided system model.}
\label{1}
\vspace{-5pt}
\end{figure}

Due to the high complexity of matrix multiplication operations, we deploy the ODE block within the transmission environment, implementing it via a distributed RISs-aided communication system as an analog ODE block during signal propagation.  The next subsection will introduce the distributed RIS-aided communication systems in greater detail.}

\subsection{{ System Model}}

 In this paper, a distributed RISs-aided communication system is considered to complete the analog Air-ODE block, where multiple-input and single-output (MISO) transceiver pairs cooperate to complete an inference task, as shown in Fig. \ref{1}. The transmitter and receiver are equipped with $N_t$ and single antennas, respectively. The relay is equipped with one receiving antenna and $N_r$ transmitting antennas to achieve data aggregation and direct the signal to another group of RISs. {  The multi-transmitter system enables each antenna to precisely delay its signal by exactly one sample relative to the previous one, ensuring consistent synchronization across varied path distances. Meanwhile, using software-defined radios enhanced with orthogonal gold sequences can accurately detect and correct symbol misalignments to achieve robust synchronization and coordination \cite{sanchez2022airnn}.
 } 
 
 In each group, there are $K$ RISs in one cluster and the number of reflective elements of each RIS is $M$. In order to clearly
  illustrate the mapping from the ODE block and the distributed RISs-aided communication system, the roles of some key system parameters are listed as follows
  
\begin{itemize}
    \item Each RIS represents a complex-valued convolution ($\mathbb{C}$Conv) layer and the number of RISs $K$ in each group is the kernel size of the convolution layer.
    \item The three groups of distributed RISs deployed in the wireless environment correspond to the three $\mathbb{C}$Conv layers in the ODE block in Fig. \ref{ODE}.
    \item  The number of directional antennas at the transmitter is $N_t = 2K + 2$ and these antennas send different signals directly to specific directions via the sophisticated antenna design.
    \item The number of transmitting antennas at the relay is $N_r = K$ and these antennas send their different signals to the corresponding RISs in group 2.
     \item  The direct links between transceivers $d_1$ and $d_2$ are considered as the residual connection.
\end{itemize}

{  As illustrated in Fig. \ref{1}, 
 the cascaded channels consist of the link \(\boldsymbol{h}_{pk} \in \mathbb{C}^{M \times 1} \) from the transmitter/relay to the \(k\)-th RIS  in the \(p\)-th group, and the link \(\boldsymbol{g}_{pk}^{T} \in \mathbb{C}^{1 \times M}\) from this RIS to the relay/receiver. Here, we assume the CSI of all the links is perfectly known,
 the receiver aggregates the features uploaded by transmitters to infer the reconstruction result and semantic tagging result of each image.  Therefore, the received signal at the relay $y_1$ and the receiver $y_2$ can be given as }
\begin{align}
     {y}_1 &=  \sum_{k = 1}^K  \underbrace{\boldsymbol{g}^T_{1k}  \boldsymbol{\Phi}_{1k}  \boldsymbol{h}_{1k}  {x}_{1k}}_{\text{Cascaded links via group 1}} + \underbrace{{d}_{1} {x}_{1}}_{\text{Direct link}} +  {n}_1, \\
     {y}_2 &=  \sum_{k = 1}^K  \underbrace{  \sum_{p = 2}^{3} \boldsymbol{g}^T_{pk}  \boldsymbol{\Phi}_{pk}  \boldsymbol{h}_{pk}  {x}_{pk}}_{\text{Cascaded links via group 2 \& 3 }}  + \underbrace{{d}_{2}{x}_{2}}_{\text{Direct link}}    + {n}_2,
     \label{received_l}
\end{align}
where the signal transmitted from each antenna directed to the $k$-th RIS in the $p$-th group is represented as ${x}_{pk}$. In addition, the signal directed to the relay and the receiver is denoted as $x_1$ and $x_2$, respectively. All the transmitted signals are subject to individual power constraints $\mathbb{E} [|| {x}_{i}||^2]  \leq P$, where $P$ represents the maximum power constraint at each antenna. ${n}_j$ is the addictive white Gaussian noise with $n_j \sim \mathcal{CN}(0, \delta_n^2)$. 

{ 
 For the \(k\)-th RIS in the \(p\)-th group, the diagonal matrix \(\boldsymbol{\Phi}_{pk}\) utilizes constant modulus and 1-bit quantization for discrete finite phase configurations. This allows for \(N_c = 2^M\) unique phase combinations, denoted as \(\{\boldsymbol{\Phi}_{\rm{ava}}\} = \{\boldsymbol{\Phi}_1, \ldots, \boldsymbol{\Phi}_{N_c}\}\). Each matrix \(\boldsymbol{\Phi}_{n}\) is defined as \(\mathrm{diag} \{ e^{j\phi_{n}^{1}}, \ldots, e^{j\phi_{n}^{M}} \}\), with possible phase values of \(\phi_0 = 0\) or \(\phi_1 = \pi/4\) for each element.
 } 
Based on the above expressions of received signals, the signal-to-noise ratio (SNR) can be given as 
\begin{align}
{\rm{SNR}}_{1} &=  \frac{\left|    \sum_{k = 1}^K   \boldsymbol{g}^T_{1k}  \boldsymbol{\Phi}_{1k}  \boldsymbol{h}_{1k}  {x}_{1k} + {d}_{1} {x}_{1} \right|^{2}}{ \sigma_1^{2}}, \\
{\rm{SNR}}_{2} &=  \frac{\left| \sum_{k = 1}^K \sum_{p = 2}^{P}   \boldsymbol{g}^T_{pk}  \boldsymbol{\Phi}_{pk}  \boldsymbol{h}_{pk}  {x}_{pk}  + {d}_{2}{x}_{2}   \right|^{2}}{  \sigma_2^{2}}.
\end{align}

\section{ Complex-valued ODE-inspired Neural Network}

{ 
In this section, we provide a detailed overview of the specific network framework and the training process used in our ODE-inspired neural network. This network, designed for dual task-oriented semantic communications, includes the encoder block, ODE block, and two decoder blocks, each tailored for specific functions. We will delve into how these components are integrated into a cohesive system and describe their individual contributions to the network's performance.

Additionally, we will outline a two-stage training process that seamlessly combines image reconstruction and semantic tagging into a unified end-to-end training model. This process is crucial for optimizing network performance across both tasks simultaneously. Each step of this training methodology will be explained in detail, illustrating how the components work together to achieve efficient learning and accurate results.
}

\subsection{The Overall Structure of the ODE-inspired Network}

As shown in Fig. \ref{digital}, the digital ODE-inspired network can be divided into three main blocks, including the encoder block, the ODE block, and the decoder block.

\subsubsection{The Encoder Block}

First, the encoder block at the transmitter compresses low-dimension features from the original complex-valued images $\boldsymbol{S}_j \in \mathbb{C}^{A \times A}$ to reduce bandwidth requirements and energy consumption. To be more specific, the image $\boldsymbol{S}_j $
is fed into an encoder block $\Omega_{\boldsymbol{\omega}}^{(\rm{EN})}(\cdot)$, resulting in $\boldsymbol{S}'_j \in \mathbb{c}^{B^2}$, given by 
\begin{align}
    \boldsymbol{S}'_j = \Omega_{\boldsymbol{\omega}}^{(\rm{EN})}  ( \boldsymbol{S}_j ),   
    \label{EN}
\end{align}
where $\Omega_{\boldsymbol{\omega}}^{(\rm{EN})}(\cdot)$ consisting of two cascaded $\mathbb{C}$Conv layers  followed by a complex-valued average pooling ($\mathbb{C}$AvgPool) function to downsample useful image features.
Then, the extracted feature $\boldsymbol{S}'_j$ is reshaped into a one-dimension (1D) feature vector $\boldsymbol{s}'_j \in \mathbb{C}^{1 \times C}$ with $C = B^2$.

\subsubsection{The ODE Block}
The ODE block is derived from the Runge-Kutta family, commonly used in numerical ODEs \cite{he2019ode}. As explained in Section II-B, the mapping between the Runge-Kutta method and the ODE-inspired neural network forms the basis for the ODE block in the proposed network structure. This block is employed for feature extraction and further learning and can be represented as
\begin{align}
     \boldsymbol{i}_j  = \frac{1}{2}\Psi_{\boldsymbol{\omega}}^{(2)}\left(    \Psi_{\boldsymbol{\omega}}^{(1)} \left(   \boldsymbol{s}'_j \right) + \boldsymbol{s}'_j 
     \right)  
      +\frac{1}{2} \Psi_{\boldsymbol{\omega}}^{(3)} \left(  \boldsymbol{s}'_j \right) +\boldsymbol{s}'_j,    
     \label{rk2}
\end{align}
where $\Psi_{\boldsymbol{\omega}}^{(p)}, p = 1,2,3$ denotes the quantized $\mathbb{C}$Conv (Q$\mathbb{C}$Conv) layers, consists of the cascaded $\mathbb{C}$ReLu function and the Q$\mathbb{C}$Conv  function.

 In addition, due to the constraints of limited feasible RIS phases in the Air-ODE network, convolutional weights in the ODE block have to be chosen from a specific set of achievable RIS phases. 
 This limitation influences the phase configurations available during the ODE training stage in the digital domain. Consequently, the trainable weights in the ODE block need to be quantized to align with the feasible weights provided by the RIS-engineered environment. First, the desired convolutional weights $W_{p}$ of each $\Psi_{\boldsymbol{\omega}}^{(p)}$ is defined as
\begin{align}
    W_{p} = \{w_{p1} , \cdots, w_{pk}, \cdots, w_{pK}  \},
\end{align}
\begin{figure}[ht]
\setlength{\abovecaptionskip}{0pt}
\setlength{\belowcaptionskip}{0pt}
\centering
 \includegraphics[width= 0.45\textwidth]{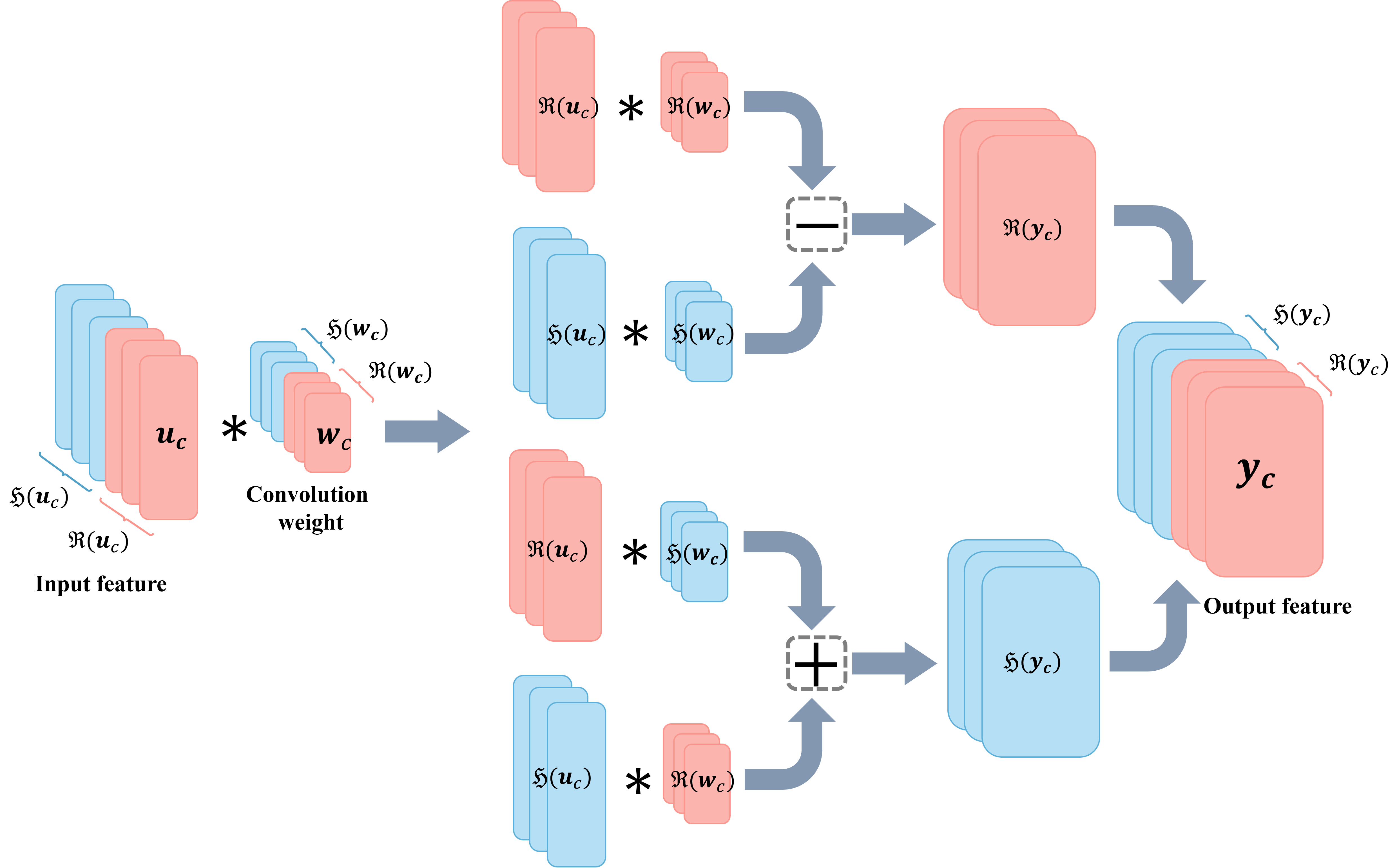}
\DeclareGraphicsExtensions.
\caption{Illustration of the complex-valued convolution operation.}
\label{CV}
\vspace{-5pt}
\end{figure} 
where $w_{pk}$ denotes the $k$-th kernel in the $\mathbb{C}$Conv layer $\Psi_{\boldsymbol{\omega}}^{(p)}$. Based on $N_c$ distinct available phase combinations, each kernel $w_{pk}$ has a corresponding set of discrete weights $ \{\tilde{h}_{pk}^{\rm{ava}}\} $ to be selected from, which is given as 
\begin{align}
  \{\tilde{h}_{pk}^{\rm{ava}}\} &= \{ \tilde{h}_{pk}^{(1)}, \cdots, \tilde{h}_{pk}^{(n)}, \cdots, \tilde{h}_{pk}^{(N_c)} \},
  \label{weight_set}
\end{align}
where $\tilde{h}_{pk}^{(n)} = \boldsymbol{g}^T_{pk}  \boldsymbol{\Phi}_{pk}^{(n)}  \boldsymbol{h}_{pk}$ denotes the $n$-th possible weight engineered by the $k$-th RIS in the $p$-th group. 

During training, we compute the Euclidean Distance (ED) from every individual weight $w_{pk}$ to the set $\{\tilde{h}_{pk}^{\rm{ava}}\}$ to find the nearest neighbor $\hat{h}_{pk}$, which is defined as 
\begin{align}
    \hat{h}_{pk}  = {\rm arg} \min_{n \in N_c} {\rm {ED}}(w_{pk}, \tilde{h}_{pk}^{(n)}),
    \label{quan}
\end{align}
where weight values $w_{pk}$ are rounded to the nearest predefined values $\hat{h}_{pk} $ for forward propagation to satisfy the hardware constraint. However, for the commonly used stochastic gradient descent (SGD) optimizer, the non-differentiable step function used in rounding impedes gradient backpropagation through the quantization function. To circumvent this, the STE approach \cite{bengio2013estimating} is employed, which approximates the derivative of the rounding function as $\boldsymbol{1}$, facilitating gradient flow.

\subsubsection{Decoder Block}

After this, the intermediate result $\boldsymbol{i}_j \in \mathbb{C}^{1 \times C}$ is reshaped as a two-dimension (2D) image, which is fed into two different decoder blocks for specific tasks. 

For semantic tagging task, the result $\boldsymbol{i}_j \in \mathbb{C}^{1 \times C}$ is fed into the decoder block $\Omega_{\boldsymbol{\omega}}^{(\text{DEC\_ST})}$ to obtain the tagging result $\boldsymbol{y}_j^{(\rm{ST})} \in \mathbb{C}^{1 \times Q}$ where $Q$ is the total number of tagging catalogues. The expression can be given as 
\begin{align}
    \boldsymbol{y}_j^{(\rm{ST})} = \Omega_{\boldsymbol{\omega}}^{(\text{DEC\_ST})}\left( \boldsymbol{i}_j \right),
    \label{DEC_ST}
\end{align}
where $\Omega_{\boldsymbol{\omega}}^{(\text{DEC\_ST})}(\cdot)$ consists of one $\mathbb{C}$Conv layer, one complex-valued Maxpooling ($\mathbb{C}$MaxPool) layer and following by a complex-valued Linear ($\mathbb{C}$Linear) layer to output the tagging result $\boldsymbol{y}_j^{(\rm{ST})} $.

For image reconstruction, the result $\boldsymbol{i}_j \in \mathbb{C}^{1 \times C}$ is fed into the decoder block $\Omega_{\boldsymbol{\omega}}^{(\text{DEC\_RI})}$ to obtain the reconstructed image $\boldsymbol{Y}_j^{(\rm{RI})} \in \mathbb{C}^{A \times A}$. The expression can be given as 
\begin{align}
    \boldsymbol{Y}_j^{(\rm{RI})} = \Omega_{\boldsymbol{\omega}}^{(\text{DEC\_RI})}\left( \boldsymbol{i}_j \right),
    \label{DEC_RI}
\end{align}
where $\Omega_{\boldsymbol{\omega}}^{(\text{DEC\_RI})}(\cdot)$ consists of one complex-valued Upsampling function ($\mathbb{C}$Upsample) layer followed by two $\mathbb{C}$Conv layers to output the reconstructed image $\boldsymbol{Y}_j^{(\rm{RI})}$.

\subsection{Model Architecture}

As introduced in the previous subsection, all functions in our proposed ODE-inspired neural network are based on complex-valued operations. In this subsection, we first detail the complex-valued convolution operations and then introduce the loss functions tailored for different tasks as well as the training algorithm.

\subsubsection{Complex-valued Operation}

First, we elaborate on the complex-valued operations employed in our network functions. These operations are essential for maintaining the integrity of the complex-valued data throughout the processing pipeline.
In the RF domain, signals are represented by complex numbers for phase and amplitude. In contrast to most neural networks which can only process real numbers and have to process complex inputs into real and imaginary parts separately, weights in distributed RISs-enabled networks are determined by limited phase configurations and are inherently coupled, which
requires the design of a new complex-valued ODE-inspired network for the subsequent deployment to distributed RISs-aided wireless environments.
To replicate traditional real-valued 2D convolution in this complex setting, we utilize a complex weight matrix $\boldsymbol{w}_c$ and corresponding input $\boldsymbol{u}_c$, which is given as 
\begin{align}
    \boldsymbol{w}_c &= \Re({\boldsymbol{w}_c})  + i\Im (\boldsymbol{w}_c), \\
    \boldsymbol{u}_c &= \Re(\boldsymbol{u}_c)  + i\Im(\boldsymbol{u}_c).
\end{align}

As illustrated in Fig. \ref{CV}, the $\mathbb{C}$Conv  is performed based on the combined convolution operations of the real and imaginary parts, where the matrix notation is given as 
\begin{align}
\left[\begin{array}{l}
\Re(\boldsymbol{w}_c  * \boldsymbol{u}_c ) \\
\Im(\boldsymbol{w}_c  * \boldsymbol{u}_c )
\end{array}\right]=\left[\begin{array}{rr}
\Re({\boldsymbol{w}}_c) & -\Im({\boldsymbol{w}}_c)  \\
\Im({\boldsymbol{w}}_c)  & \Re({\boldsymbol{w}}_c) 
\end{array}\right] *\left[\begin{array}{l}
\Re({\boldsymbol{u}}_c) \\
\Im(\boldsymbol{u}_c)
\end{array}\right].
\end{align}

Following the same principle, the $\mathbb{C}$Linear function can be  constructed as 
\begin{align}
   \boldsymbol{w}_l 
   \cdot \boldsymbol{u}_l &=\left(\Re(\boldsymbol{w}_l)  \cdot \Im({\boldsymbol{u}_l})  -\Im(\boldsymbol{w}_l)  \cdot \Im(\boldsymbol{u}_l) \right) \\ \nonumber
   &+i \left( \Im(\boldsymbol{w}_l) \cdot \Re(\boldsymbol{u}_l)  + \Re(\boldsymbol{w}_l)  \cdot \Im(\boldsymbol{u}_l)  \right),
\end{align}
where $\boldsymbol{w}_l$ and $\boldsymbol{u}_l$ denote the weight and the input of the $\mathbb{C}$Linear layer.
Similar to the $\mathbb{C}$Conv  and $\mathbb{C}$Linear functions, our network incorporates other complex-valued functions based on the same principles. For instance, the $\mathbb{C}$ReLu function is defined as
\begin{align}
    \mathbb{C}{\rm{ReLU}}(\boldsymbol{z}) = {\rm{ReLU}}\left( \Re(\boldsymbol{z}) \right) + i \cdot {\rm{ReLU}}\left( {\Im (\boldsymbol{z})}\right).
    \label{crelu}
\end{align}

Building on this framework, additional complex-valued functions are integrated into our network structure following a consistent operational philosophy. { Specifically, the Q\( \mathbb{C} \)Conv function operates similarly to \( \mathbb{C} \)Conv but with the distinction that the weights are selected from a feasible set determined by RIS phases, as outlined in Eq. (\ref{quan}). Moreover, functions such as \( \mathbb{C} \)AvgPool, \( \mathbb{C} \)MaxPool, and \( \mathbb{C} \)Upsample are defined analogously to \( \mathbb{C} \)ReLU, as specified in Eq. (\ref{crelu}). These adaptations ensure that all complex-valued operations maintain consistency with the fundamental principles of complex-valued processing within the network.}

\subsubsection{Loss Functions}

For the task of image reconstruction, the decoder block $\Omega_{\boldsymbol{\omega}}^{(\text{DEC\_ST})}(\cdot)$  is designed to recover high-quality images from the compressed features. To evaluate the reconstruction performance during training, the MSE is employed as the loss function, defined as
\begin{align}
   L^{\rm{R}}(\boldsymbol{Y}_j^{(\rm{RI})} \! \!, \boldsymbol{S}_{j}) = \frac{[\Re(\boldsymbol{Y}_j^{(\rm{RI})}) \!- \! \Re({\boldsymbol{S}_{j}})]^2 \! \!+ \!\! [\Im(\boldsymbol{Y}_j^{(\rm{RI})}) \!-\!\Im({\boldsymbol{S}_{j}})]^2}{2A^2 },
\end{align}
where $A^2$ is the total number of pixels in the image. 

In the semantic tagging task, the cross-entropy (CE) is used as the loss function to evaluate the tagging accuracy. The decoder $\Omega_{\boldsymbol{\omega}}^{(\text{DEC\_ST})}(\cdot)$ is designed to produce the tagging prediction result $\boldsymbol{y}_j^{(\rm{ST})}$. The output $\boldsymbol{y}_j^{(\rm{ST})}$ from the complex-valued neural network is first converted to its modulus value, and then the loss function can be calculated.
The CE function, based on the tagging label $\boldsymbol{l}_j$, is defined as 
 \begin{align}
      L^{\rm{S}}(|\boldsymbol{y}_{j}^{(\rm{ST})}|,\boldsymbol{l}_j ) = -\sum_{q=1}^Q  {l}_{j,q} \log\left( \frac{e^{| {y}_{j,q}^{(\rm{ST})}|}} {\sum_{q' = 1}^Q e^{ |{y}_{j,q'}^{(\rm{ST})}|}}\right).
 \end{align}
Since the aim of this network is to minimize the error of both loss functions simultaneously. Thus, the objective for joint training of the proposed ODE-inspired neural network should be
\begin{align}
\label{J_loss}
L_{\text{total}} &= \alpha L^{\rm{R}}(\boldsymbol{Y}_j^{(\rm{RI})} \! \!, \boldsymbol{S}_{j}) + \beta  L^{\rm{S}}(|\boldsymbol{y}_{j}^{(\rm{ST})}|,\boldsymbol{l}_j ) \\\nonumber
&= \alpha \cdot \frac{[\Re(\boldsymbol{Y}_j^{(\text{RI})}) - \Re(\boldsymbol{S}_j)]^2 + [\Im(\boldsymbol{Y}_j^{(\text{RI})}) - \Im(\boldsymbol{S}_j)]^2}{2A^2} \\ \nonumber
&-\sum_{q=1}^Q \beta \cdot  {l}_{j,q} \log\left( \frac{e^{| {y}_{j,q}^{(\text{ST})}|}}{\sum_{q' = 1}^Q e^{| {y}_{j,q'}^{(\text{ST})}|}} \right). 
\end{align}
Although it is common to use penalty terms $\alpha$ and $\beta$ to balance the impact of two different losses on overall performance, this method is not directly suitable for our task. 
 
First, for image reconstruction, the MSE is used to evaluate the difference of each pixel, calculated by 
$  L^{\rm{R}}(\boldsymbol{Y}_j^{(\rm{RI})} \! \!, \boldsymbol{S}_{j})$. Since MSE measures the square of pixel differences, even small pixel variations are significantly amplified, leading to substantial changes in the loss during the early stages of training. For instance, if there are small discrepancies between the predicted and the actual image pixels, the MSE will magnify these differences, resulting in large loss variations.
 
In contrast, the error for  $ L^{\rm{S}}(|\boldsymbol{y}_{j}^{(\rm{ST})}|,\boldsymbol{l}_j )$  ranges between 0 and 1, and the tagging task is relatively simpler as it only involves distinguishing discrete class labels. The loss function for the tagging task shows smaller variations and converges more easily. For example, a classification task merely needs to determine which category an image belongs to, with a lower probability of error and thus a smaller loss.

Therefore, directly using penalty terms to balance the losses would cause $\alpha$ to become very small due to the larger loss from image reconstruction and the simpler tagging task. Specifically, for a small perturbation to the value of $\alpha$, the network training would be partial to the semantic tagging task, resulting in unsatisfactory outcomes in terms of image reconstruction accuracy. To address this, we propose a two-stage fine-tuning algorithm, as outlined below.

\subsection{Fine-tuning Algorithm}

Given the differences in loss functions, we first train the network for image reconstruction, then freeze most of it for fine-tuning both tasks, optimizing results as detailed in Algorithm \ref{alg:Framwork1}.

\begin{algorithm}[htb]
\caption{ The Training Process of the Complex-valued ODE-inspired Neural Network  }
\label{alg:Framwork1}
\begin{algorithmic}[1] 
\REQUIRE ~~\\ 
    Limited feasible weights enabled by distributed RISs $ \{\tilde{h}_{pk}^{\rm{ava}}\}, \forall k = 1,\cdots, K$ and $p = 1,2,3$. \\
    The training and testing samples $\mathcal{J} = \{({\boldsymbol{S}_{j}}, \boldsymbol{l}_{j})\}_{j = 1}^{J}$. 
\ENSURE ~~\\ 
    The constructed complex-valued image $\{\boldsymbol{Y}_j^{(\rm{RI})}\}_{j = 1}^J$.\\
    The semantic tagging of the image $\{ \boldsymbol{y}_j^{(\rm{ST})} \}_{j = 1}^J$.
    \STATE  \textbf{Stage 1: Training only focus on the image construction:}
    \STATE  Freeze the semantic tagging decoder $\Omega_{\boldsymbol{\omega}}^{(\text{DEC\_ST})}(\cdot)$;
    \FOR {epoch $  = 1, \cdots, N_{e_1} $}
    \STATE Extract and compress the feature  $\boldsymbol{s}'_j$  from the original image ${\boldsymbol{S}_{j}}$ by (\ref{EN});
    \STATE Feed $\boldsymbol{s}'_j$  to the ODE block for further learning by $(\ref{rk2})$;
    \STATE Generate $\boldsymbol{Y}_j^{(\rm{RI})}$ via $\Omega_{\boldsymbol{\omega}}^{(\text{DEC\_RI})}(\cdot)$ by (\ref{DEC_RI});
    \STATE Calculate the MSE loss and update the weight through the STE approach by (\ref{quan});
    \ENDFOR
    \STATE \textbf{Stage 2: Joint fine-tuning for the image reconstruction and semantic tagging: }
     \STATE Freeze the encoder block $\Omega_{\boldsymbol{\omega}}^{(\rm{EN})}(\cdot)$  and ODE block based on the trained model in stage 1; 
     \STATE Unfreeze two decoder blocks for image reconstruction $\Omega_{\boldsymbol{\omega}}^{(\text{DEC\_RI})}(\cdot)$ and semantic tagging $\Omega_{\boldsymbol{\omega}}^{(\text{DEC\_ST})}(\cdot)$;
    \FOR {epoch $  = 1, \cdots, N_{e_2} $}
    \STATE  Same operations as in steps 3 $\sim$ 5 from Stage 1;
    \STATE Generate the outputs $\boldsymbol{Y}_j^{(\rm{RI})}$ and  $\boldsymbol{y}_j^{(\rm{ST})}$ by (\ref{DEC_RI}) and (\ref{DEC_ST});
    \STATE Calculate the joint loss using (\ref{J_loss}) and update the weights in both decoder blocks;
    \ENDFOR
\RETURN $\{\boldsymbol{Y}_j^{(\rm{RI})}, \boldsymbol{y}_j^{(\rm{ST})}\}_{j = 1}^J$ and $\hat{h}_{pk}, \forall k = 1,\cdots, K$ and $p =1,2,3$. 
\end{algorithmic}
\end{algorithm}

Specifically, in Stage 1, the decoder block responsible for semantic tagging is frozen. This ensures that the end-to-end training process focuses exclusively on the image reconstruction task. During this stage, the ODE block is updated using the STE approach, with weights selected from a set of feasible options engineered by distributed RISs. 

In Stage 2, the encoder block and the ODE block of the network are frozen to retain the learned features and transformations from Stage 1. Meanwhile, the two decoder blocks, responsible for image reconstruction and semantic tagging, are unfrozen. The training during this stage utilizes the joint loss function (\ref{J_loss}), 
 which balances both tasks. By fine-tuning these decoder blocks, the network refines its predictions to produce the final reconstructed image and accurate semantic tagging results.
\begin{figure}[ht]
\setlength{\abovecaptionskip}{0pt}
\setlength{\belowcaptionskip}{0pt}
\centering
\includegraphics[width= 0.35\textwidth]{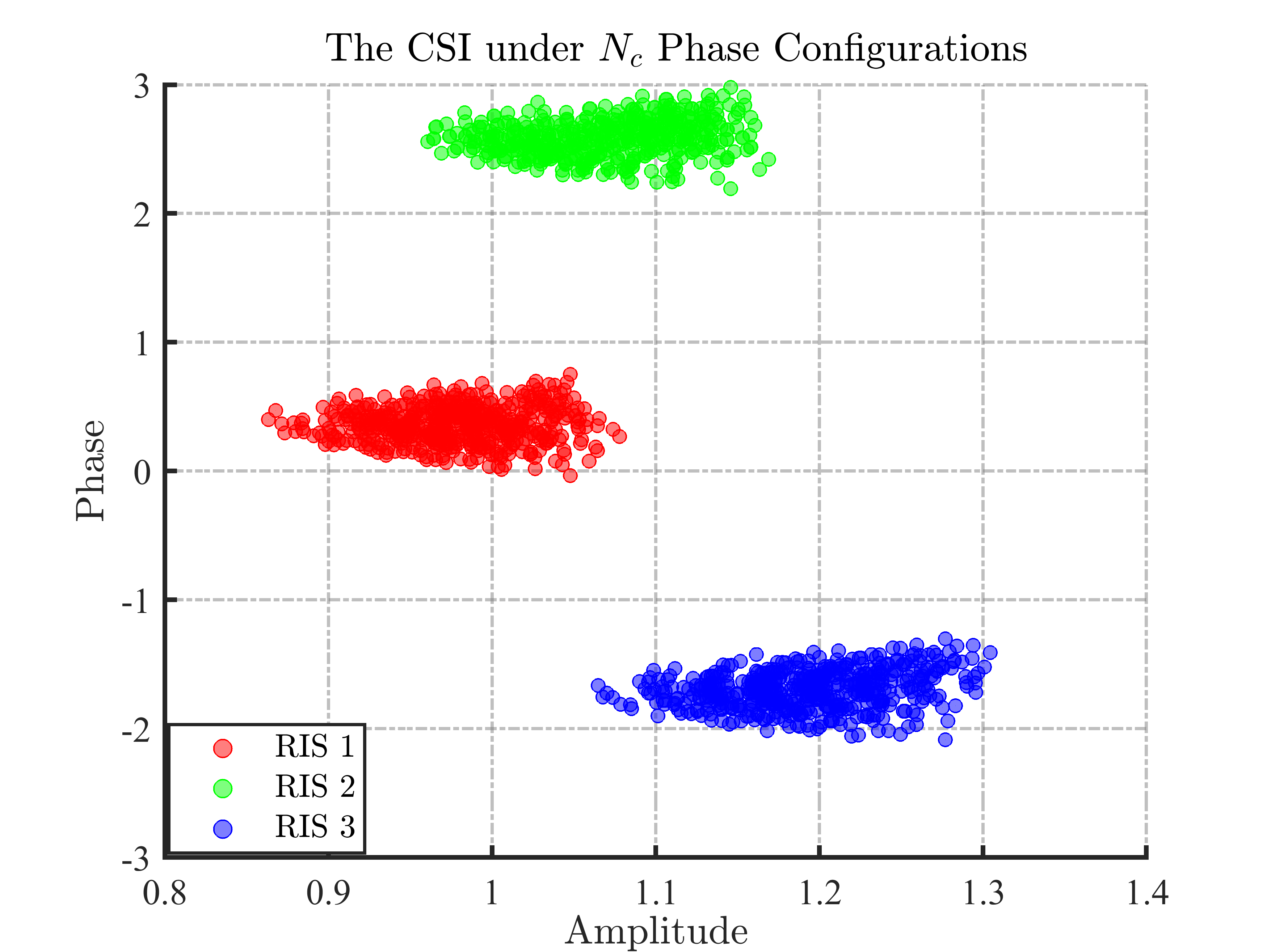}
\DeclareGraphicsExtensions.
\caption{CSI for each RIS.}
\vspace{-5pt}
\label{csi_a}
\end{figure} 
\begin{figure}[ht]
\setlength{\abovecaptionskip}{0pt}
\setlength{\belowcaptionskip}{0pt}
\centering
\includegraphics[width= 0.35\textwidth]{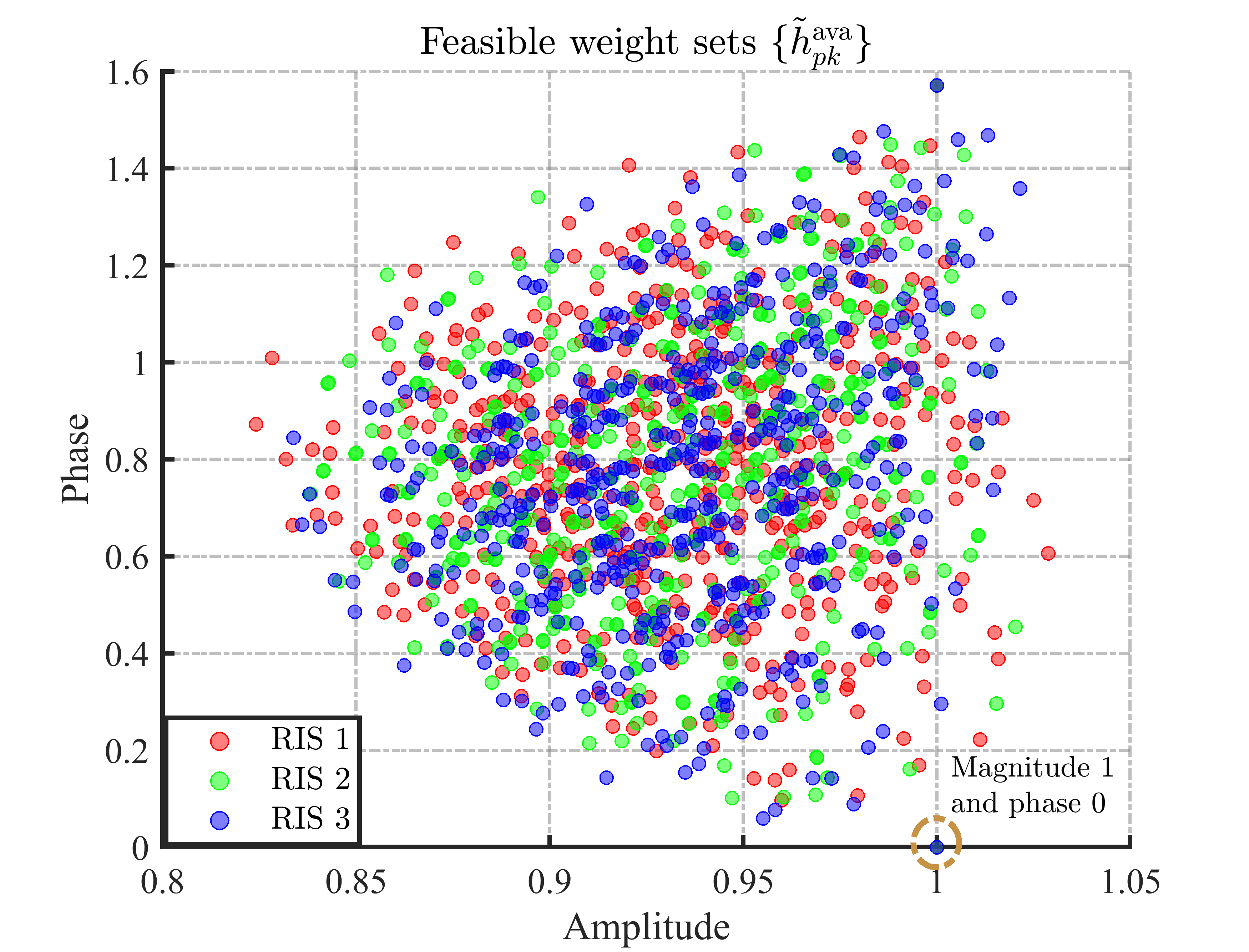}
\DeclareGraphicsExtensions.
\caption{Feasible weights for each RIS.}
\vspace{-5pt}
\label{csi_b}
\end{figure} 

\section{The Deployment of the Analog Air-ODE Block via Distributed RISs}

After the end-to-end training, the ODE block in the complex-valued ODE-inspired neural networks will be deployed in the wireless environment, while the encoder and decoder blocks are deployed at the transmitter and receiver, respectively. Consequently, the ODE block can be implemented through signal transmission via multiple groups of distributed RISs, achieving the Air-ODE block. In this section, we will detail how to construct the wireless communication environment for the trained ODE-inspired neural network.

\subsection{Precoding for Channel Tracing}
 
Due to the dynamic nature of wireless channels, the CSI of each communication link may fluctuate over time, making it impractical to retrain the model to adapt to these time-varying channels. To circumvent the need for retraining the model under such conditions, Air-ODE employs channel tracking and correction techniques. These methods ensure that the weights of the convolution layer, as determined by the RIS configuration, remain valid even under new channel conditions induced by slow fading \cite{sanchez2022airnn}.

\begin{figure}[ht]
\setlength{\abovecaptionskip}{0pt}
\setlength{\belowcaptionskip}{0pt}
\centering
\includegraphics[width= 0.48\textwidth]{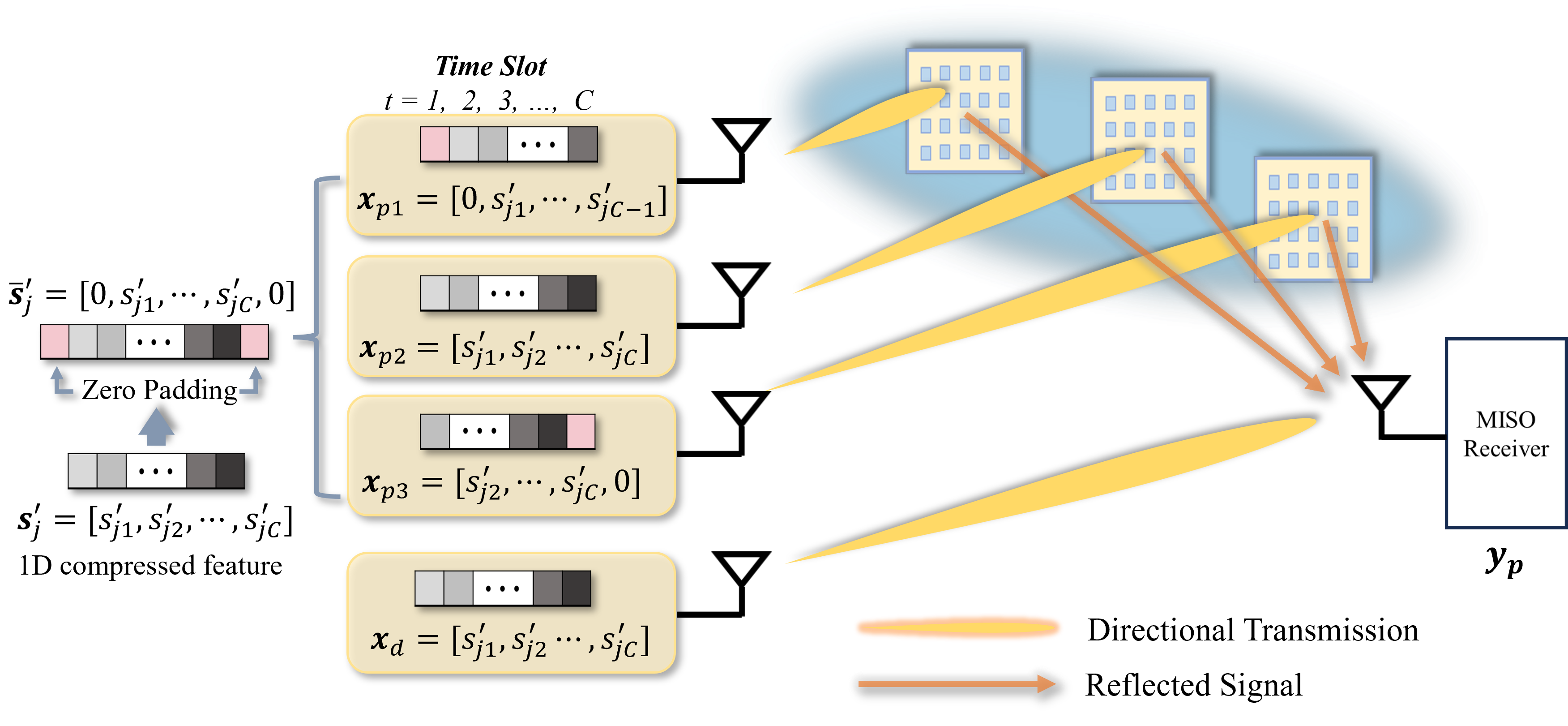}
\DeclareGraphicsExtensions.
\caption{Padding and analog convolution within a single group of RISs.}
\label{zp}
\vspace{-5pt}
\end{figure}

For each RIS, the first phase configuration $\boldsymbol{\Phi}_{1}$ from the set $\{\boldsymbol{\Phi}_{\rm{ava}}\}$ is selected as the baseline phase configuration. The corresponding channel response is given by $\dot{h}_{pk}^{(1)} = \boldsymbol{g}_{pk}^{T} \boldsymbol{\Phi}_{1}\boldsymbol{h}_{pk}$.
Then, the transmitter 
obtain the estimated channel vector $\dot{h}_{pk}^{(1)}$ and the precoding design can be presented as $v_{pk} = 1 / \dot{h}_{pk}^{(1)} $, to equalize the channel transformation for a given RIS configuration in $\{\boldsymbol{\Phi}_{\rm{ava}}\}$, which ensures that there is always a weight with magnitude 1 and phase 0 within the set of feasible weights. Then, we use $v_{pk}$ to capture the phase rotation of the other phase configurations in the set $\{\boldsymbol{\Phi}_{\rm{ava}}\}$, and the expression is given by

\begin{align}
    \cos(\alpha_{pk}^{(n)}) =  {\rm Re}\{(h_{pk}^{(n)})^H v_{pk}\} / (||h_{pk}^{(n)} ||\cdot||v_{pk}||).
    \label{phase_rotation}
\end{align}

By applying this rotation to all available phase configurations in $\{\boldsymbol{\Phi}_{\rm{ava}}\}$, i.e., $\tilde{h}_{pk}^{(n)} = h_{pk}^{(n)}e^{j\alpha_{pk}^{(n)}}$, the feasible weight set $\{\tilde{h}_{pk}^{\rm{ava}}\}$ enabled by $\{\boldsymbol{\Phi}_{\rm{ava}}\}$  can be given as (\ref{weight_set}).

As shown in Fig. \ref{csi_a}, we illustrate the CSI of cascade links associated with three different RISs within the same group. It can be observed that, due to the different deployment locations of the RISs, the CSI distributions related to the three RISs form distinct clusters within the same group. Each cluster corresponds to $N_c = 512 $ possible phase configurations. Consequently, the resulting weights may not exactly match those required for a digital convolution operation, and it is impractical to retrain the neural network whenever the channel changes.

After rotation, the three different sets of RIS-related CSI distributions consistently have a weight of magnitude of 1 and a phase of 0, as shown within the yellow circle in Fig. \ref{csi_b}. Additionally, unlike the three scattered clusters in Fig. \ref{csi_a}, all viable weights in Fig. \ref{csi_b} are distributed in the same area, thus alleviating the retraining issues caused by varying locations and time-varying channels.

\subsection{ The Deployment of Analog Air-ODE}

In a sensing stage, the transmitter acquires a complex-valued image from the local dataset $\mathcal{J} = \{{\boldsymbol{S}_{j}}, \boldsymbol{l}_j\}_{j = 1}^{J}$ consisting of $J = |\mathcal{J}|$ training data samples, where $\boldsymbol{S}_{j} \in \mathbb{C}^{A \times A} $ and $\boldsymbol{l}_j$ are the sample $j$ and the corresponding tagging label, respectively. 
 Then, the transmitter extracts the compressed feature $\boldsymbol{s}_{j}\in \mathbb{C}^{ 1 \times C}$ from ${\boldsymbol{S}_{j}}$ as the signal. As the entries in each local gradient may vary significantly, the devices need to transform the gradients into normalized signal vectors to facilitate pre-equalization and power control \cite{zhu2019broadband}.  
To this end, each device normalizes the features into a normalized symbol vector $\boldsymbol{s}_j$ with zero mean and unit variance.

In the context of RISs that passively reflect signals and only implement multiplication, additional steps are necessary to complete the convolution operation. One such step is padding before directional transmission at each antenna to preserve the original image information, ensuring that the feature map size after convolution remains consistent with the input image. This approach simulates digital convolution operations and necessitates signal adjustment before transmission. To prevent data distortion, the edges of $\boldsymbol{s}_j$ are zero-padded, ensuring interference-free data.
\begin{algorithm}[htbp]
\caption{ Deployment of the Distributed RISs-aided Communication System}
\label{alg:offline_deployment}
\begin{algorithmic}[1]
\REQUIRE ~~\\
The optimal weight enabled by distributed RISs \( \hat{h}_{pk} \), \(\forall k = 1, \ldots, K\) and \( p = 1, 2, 3 \).\\
The training and testing samples \( \mathcal{J} = \{(\boldsymbol{S}_j, \boldsymbol{l}_j)\}_{j=1}^J \).
\ENSURE ~~\\
The constructed complex-valued image \( \{\boldsymbol{Y}_j^{(\text{RI})}\}_{j=1}^J \).\\
The semantic tagging of the image \( \{\boldsymbol{y}_j^{(\text{ST})}\}_{j=1}^J \).
\end{algorithmic}
\hspace{0.5em}\textbf{Procedure:}
\begin{algorithmic}[1]
\STATE Estimate the CSI \( \hat{h}_{pk}^{(1)} \) and use the precoding value to rotate all the available phases for channel tracking by (22).
\STATE Zero-pad the transmitted signal at each antenna.
\STATE Configure the RISs according to the indexes of \( \hat{h}_{pk} \).
\STATE Transmit all the signals in sequential time slots to the receiver by (25).
\STATE Feed the received signals to the decoder blocks \( \Omega_{\boldsymbol{\omega}}^{(\text{DEC\_RI})} \) and \( \Omega_{\boldsymbol{\omega}}^{(\text{DEC\_ST})} \).
\RETURN \( \{\boldsymbol{Y}_j^{(\text{RI})}, \boldsymbol{y}_j^{(\text{ST})}\}_{j=1}^J \).
\end{algorithmic}
\end{algorithm}
Fig. \ref{zp} illustrates a simple analog residual block, where three RISs correspond to a $1\times3$ $\mathbb{C}$Conv function for easier understanding,  and the direct link represents the residual connection of the residual block. For the 1D compressed feature $\boldsymbol{s}_j$,  containing $C$ elements,  zero-padding is applied to both ends to match the original image size, preventing loss of feature information. This results in $\bar{\boldsymbol{s}}'_{j} \in \mathbb{C}^{1\times(C+2)} =  [0,\bar{{s}}'_{j1},\cdots, \bar{{s}}'_{ji},\cdots \bar{{s}}'_{jC},0]$\footnote{The amount of zero-padding applied on each side of the vector in the corresponding digital ODE-based network depends on the convolution kernel size. For a kernel size of 
$ 1 \times 3$, one zero is added to each edge of the vector.}. This padding strategy is essential for preserving data integrity and ensuring that the distributed RISs-aided communication systems can effectively emulate digital $\mathbb{C}$Conv processes without introducing distortions.

The padded vector $\bar{\boldsymbol{s}}'_{j}$ is divided into three segments based on the convolution kernel size.
To ensure accurate channel tracking, each segment of the padded vector $\bar{\boldsymbol{s}}'_{j}$ is multiplied by 
 $v_{pk}$, resulting in $\boldsymbol{x}_{pk} \in \mathbb{C}^{1 \times C}$.
Each vector $\boldsymbol{x}_{pk}$ is then transmitted to the specific $k$-th RIS via a directional antenna, with each element in the vector transmitted over $C$ sequential time slots. Consequently, the aggregated features ${y}_{p}^{(t)}$ received  at a certain time slot $t$ can be represented as
\begin{align}
    {y}_{p}^{(t)} = \underbrace{\sum_{k = 1}^{3} \hat{h}_{pk}^{(t)} x_{pk}^{(t)}}_{\text{ $1\times 3$ Convolution}}  + \quad x_d^{(t)},
\end{align}
 where $\hat{h}_{pk}^{(t)} $ represents the weight engineered by the $k$-th RIS in the $p$-th group at time slot $t$. Thus, the received feature vector is denoted as $\boldsymbol{y}_{p} = [y_p^{(1)}, \cdots, y_p^{(t)}, \cdots, y_p^{(C)}]$.

To achieve convolution operations via distributed RISs, it is imperative to maintain constant inter-path time arrivals from consecutive RIS paths. This necessitates ensuring a precise match between these inter-path time arrivals and the communication symbol time. Accordingly, a multi-antenna system is deployed, where each antenna transmits signals with a time delay of exactly one sample relative to the next, thereby maintaining equal spacing between arriving signals \cite{sanchez2022airnn}.
Based on this transmission scheme, similar to the precoding design for cascaded links, the signals directed to the direct links $d_1$ and $d_2$ can be presented as $v_1 = 1/d_1$ and  $v_2 = 1/d_2$, respectively. Therefore, the signal  received through Air-ODE enabled by distributed RISs at time slot $t$ can be rewritten as 
 \begin{align}
    y^{(t)} & = \sum_{k_3 = 1}^3  \hat{h}_{3k_3}^{(t)} x_{3k_3}^{(t)} + x_2^{(t)} + {n}_2^{(t)} \\ \nonumber
     & \! +\!  \sum_{k_2 = 1}^3  \frac{\hat{h}_{2k_2}^{(t)} }{2}    \!\left(  {\sum_{k_1 = 1}^{3} \hat{h}_{1k_1}^{(t)} x_{1k_1}^{(t)} + x_1^{(t)} + n_1^{(t)} } \right).
     \label{aggre_fea}
 \end{align}
After receiving the output feature $\boldsymbol{y} = [y^{(1)},\cdots,y^{(C)}]$, the receiver can infer the reconstructed image $\boldsymbol{Y}_j^{(\rm{RI})}$ and semantic tagging result $\boldsymbol{y}_j^{(\rm{ST})}$ through the decoder  $\Omega_{\boldsymbol{\omega}}^{(\text{DEC\_RI})}$ and $\Omega_{\boldsymbol{\omega}}^{(\text{DEC\_ST})}$, respectively. The above processing steps are summarized in Algorithm 2.
 
    

\section{Numerical Results}
\subsection{Parameters Setup}
Throughout the simulation,  each RIS group consists of three RISs, and the number of reflective elements $M$ at each RIS is nine with dimensions $M_x = M_y = 3$. Based on 1-bit quantization, the feasible phase configurations for each element are $0^\circ$ and $45^\circ$ \cite{sanchez2022airnn}, resulting in a total of $N_c = 2^9$ feasible phase combinations.
We assume that the direct links ${d}_1$ and ${d}_2$ follow Rayleigh fading, while   cascaded channels $\boldsymbol{g}_{pk}$ and $\boldsymbol{h}_{pk}$ follow the Rician fading. Similar to the configuration \cite{guo2020weighted}, the antenna elements are assumed to form a half-wavelength spaced uniform linear array configuration at both the transceivers and RISs. Therefore, the cascaded channels $\boldsymbol{g}_{pk}$ and $\boldsymbol{h}_{pk}$ are modeled by
\begin{align}
    \boldsymbol{g}_{pk}  \! &= \! \sqrt{L_{1,pk} }\left( \sqrt{\frac{\rho}{1\!+\!\rho}} \boldsymbol{a}_M(\vartheta_
{pk})  \!+ \!\sqrt{\frac{1}{1+\rho}} \overline{\boldsymbol{g}}_{pk}   \right),\\
    \boldsymbol{h}_{pk}  &= \sqrt{L_{2,pk}} \left( \sqrt{\frac{\rho}{1+\rho}} \boldsymbol{a}_M(\varsigma_{pk})   + \sqrt{\frac{1}{1+\rho}} \overline{\boldsymbol{h}}_{pk}  \right),
\end{align}
where $L_{1,pk}$ and $L_{2,pk}$ represent the path losses for the cascaded channels and are sampled from the complex normal distribution $\mathcal{CN}\left(0,10^{-0.1 \mathrm{PL}(r)}\right)$. The path loss function $\mathrm{PL}(r)= \varrho_a + 10 \varrho_b \log(r)$, with parameters $\varrho_a  = 35.6 $ and $\varrho_{\mathrm{b}}=2.2$. Additionally, $\rho = 10$ represents the Rician factor and $\boldsymbol{a}$ the steering vector. Angular parameters are denoted by $\vartheta_{pk},  \varsigma_{pk}$, while the non-line-of-sight (NLoS) components $\overline{\boldsymbol{g}}_{pk}$ and $\overline{\boldsymbol{h}}_{pk}$ follow the distribution $\mathcal{CN}(0,1)$. We tested the Air-ODE network with the complex-valued Fashion-MINST dataset \cite{lin2018all}, which utilizes amplitude and phase channels for data representation. The dataset contains 60,000 samples, with 54,000 used for training and 6,000 for validation. { 
Regarding the training process, the total training encompasses 300 epochs, divided into two stages. The initial stage comprises 200 epochs dedicated solely to image construction, followed by a second stage of 100 epochs focused on joint fine-tuning of image reconstruction and semantic tagging. A learning rate of $1 \times 10^{-4}$ is employed, with a batch size of 64. The optimization is performed using the Adam optimizer on an NVIDIA RTX 4070 Ti GPU.
 }

Performance was benchmarked against a digital ODE-inspired neural network, and the effectiveness of the Air-ODE structure was further assessed in scenarios with random phase configurations and without the Air-ODE structure.

\subsection{Network Structure}
The digital ODE-inspired neural network is structured into three main blocks: the encoder block, the Air-ODE block, and the decoder block. The detailed configuration of each block is as follows
\begin{itemize}
\item {\textbf{Encoder block:}} 
 The input undergoes two $\mathbb{C}$Conv layers with 3$\times$3 kernels, each followed by a $\mathbb{C}$ReLU function. The output is then downsampled using a  $\mathbb{C}$AvgPool layer with a kernel size of 2 and a stride of 2, which is followed by another $\mathbb{C}$ReLU, and one $\mathbb{C}$BatchNorm.

\item {\textbf{Air-ODE block:}} 
This block comprises three quantized  $\mathbb{C}$Conv layers with 1x3 kernels. Unlike the convolution operations in the encoder block, these quantized $\mathbb{C}$Conv layers are optimized using the feasible weight sets enabled by the RISs. These layers are responsible for the Air-ODE block, with their outputs contributing to the final combined feature map.

\item {\textbf{Decoder block:}}
The feature map is upsampled using a $\mathbb{C}$Upsample with a scale factor of 2, followed by two 3x3 $\mathbb{C}$Conv with $\mathbb{C}$ReLU activations to obtain the reconstructed image. For semantic tasks, the feature map passes through another 3x3 $\mathbb{C}$Conv layer, a  $\mathbb{C}$MaxPool layer with a kernel size of 4 and a stride of 4, then flattened and processed by a  $\mathbb{C}$Linear layer. The output magnitudes are computed for the final tagging result.

\end{itemize}
 \begin{figure*}[ht]
\centering
\includegraphics[scale=.8]{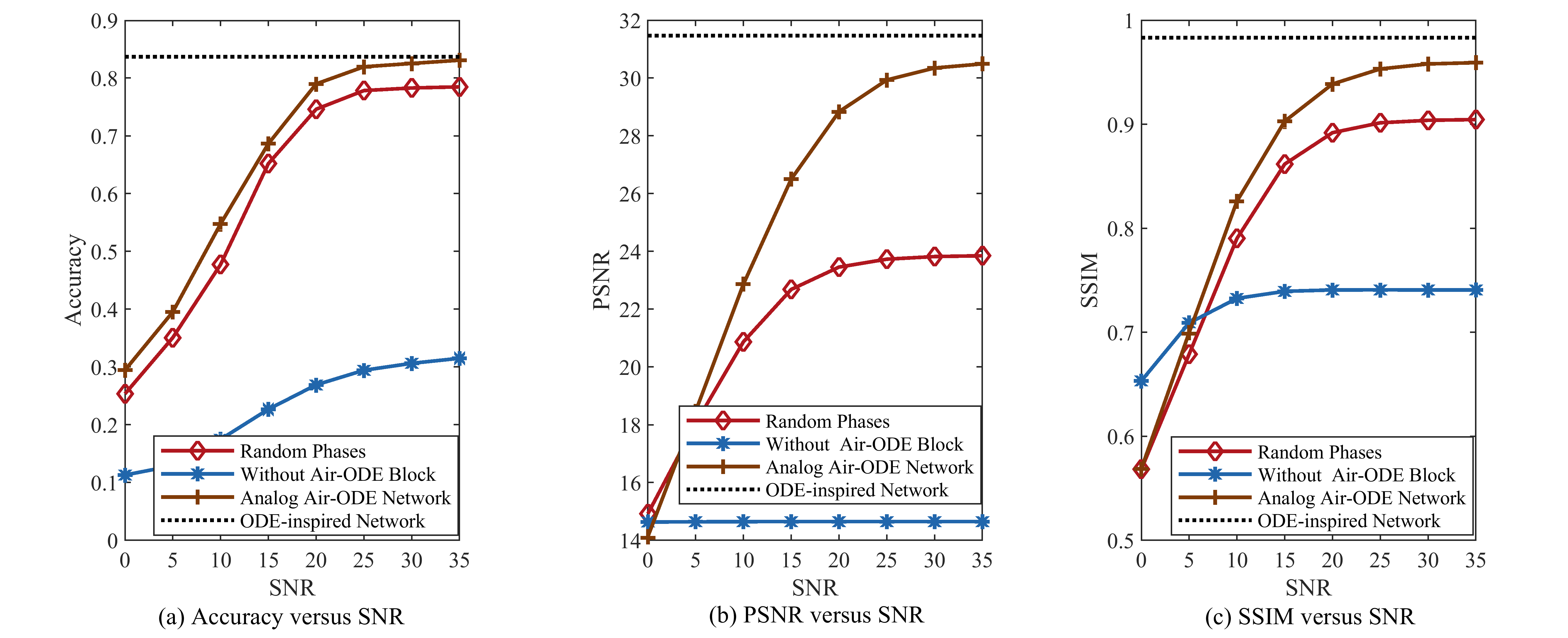}
\caption{Performance versus SNR. }
\label{SNR}
\end{figure*}
\begin{figure*}[ht]
\centering
\includegraphics[scale=.45]{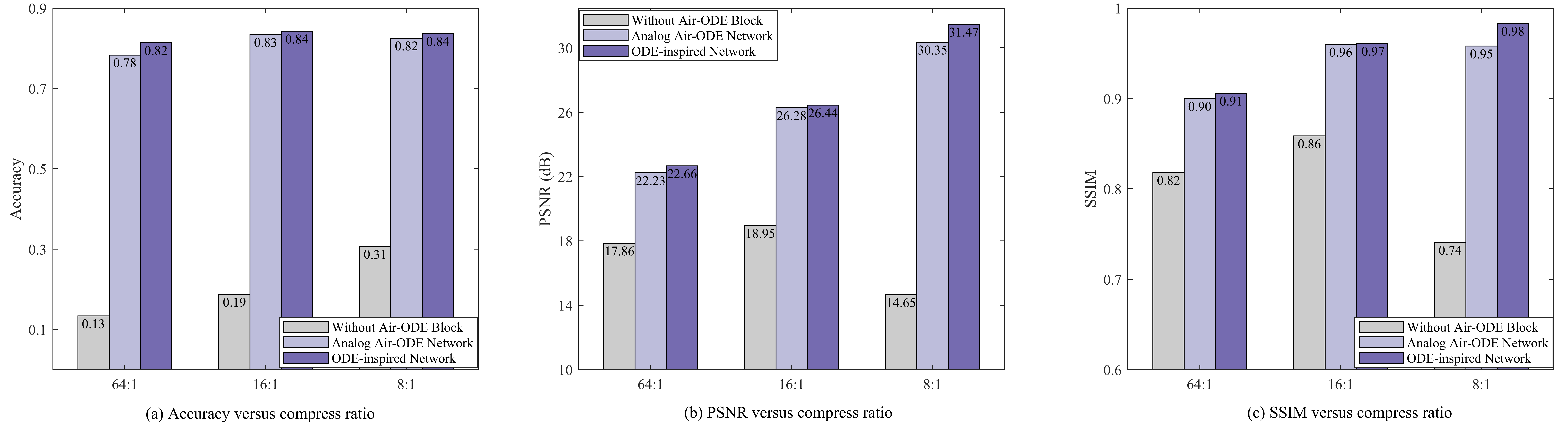}
\caption{Performance versus compress ratio. }
\label{compress}
\end{figure*}
\subsection{Performance Metrics}

The semantic tagging accuracy is used to evaluate the successful prediction rate output by the ODE-inspired neural network. When evaluating the similarity between an image output and its true example, various metrics can be utilized. Two widely used metrics are the PSNR and the  SSIM, which assess the similarity of two images in different ways. The measurements are listed as follows

\begin{itemize}
\item  Consider a tagging scenario with \(Q\) distinct tags, the overall accuracy for this model can then be mathematically expressed as
\begin{align}
    \text{Accuracy} = \frac{\sum_{q=1}^{A} \Lambda_{qq}}{\sum_{q=1}^{Q} \sum_{q'=1}^{C} \Lambda_{qq'}},
\end{align}
where $ \sum_{q=1}^{Q} \Lambda_{qq'} $ encapsulates the total number of instances correctly predicted across all classes in the confusion matrix $\Lambda$.
\begin{table}[!h]
\caption{\textbf{PERFORMANCE COMPARISON UNDER DIFFERENT SCENARIOS }}
 \centering
 \scalebox{1}
 {
\begin{tabular}{c c c c}
\hline
\specialrule{0em}{1pt}{1pt}
\hline
 \diagbox{Scenarios}{Performance}&  PSNR&   SSIM & Accuracy \\
 \hline
ODE-inspired Network&  31.4664 & 0.9832 & 83.68$\%$ \\
\hline
{ DeepJSCC}&  31.4205 &  { 0.9764} & { $82.63 \%$} \\
\hline
Proposed analog Air-ODE &  30.3476 & 0.9580 & $82.52\%$ \\
\hline
\specialrule{0em}{1pt}{1pt}
\hline
\end{tabular}}
\end{table}
    \item The PSNR focuses on the details of the images, similar to the  MSE \cite{hore2010image}, which evaluates individual pixels, and the expression is defined as
\begin{align}
{\rm {PSNR}} =10 \log _{10}\left(\frac{{\rm{MAX_I}}^2}{\rm {MSE}}\right),
\end{align}
where ${\rm{MAX_I}}$ is the dynamic range of the pixel values in the image.
\item  The SSIM focuses on the overall structure of the image and is defined as \cite{hore2010image}
\begin{align}
\operatorname{SSIM}(\boldsymbol{Y}_j^{(\rm{RI})} \! \!, \boldsymbol{S}_{j})=\frac{\left(2 \mu_{\boldsymbol{Y}} \mu_{\boldsymbol{S}}+c_1\right)\left(2 \sigma_{{\boldsymbol{Y}}{\boldsymbol{S}}}+c_2\right)}{\left(\mu_{\boldsymbol{Y}}^2+\mu_{\boldsymbol{S}}^2+c_1\right)\left(\sigma_{\boldsymbol{Y}}^2+\sigma_{\boldsymbol{S}}^2+c_2\right)},
\end{align}
where $\mu_{\boldsymbol{Y}}$ and $\mu_{\boldsymbol{S}}$ are the mean pixel values of the two images $\boldsymbol{Y}_j^{(\rm{RI})} $ and $\boldsymbol{S}_{j}$, $\sigma_{\boldsymbol{Y}}$ and $\sigma_{\boldsymbol{S}}$ are the standard deviations of the pixel values of the two images $\boldsymbol{Y}_j^{(\rm{RI})} $ and $\boldsymbol{S}_{j}$, $c_1$ and $c_2$ are constants defined as $\left(K_1 {\rm{MAX_I}}\right)^2$ and $\left(K_2 {\rm{MAX_I}}\right)^2$, respectively. $K_1$ and $K_2$ are two small constants $\left(K_1, K_2 \ll 1\right)$.

\end{itemize}

\subsection{Simulation Results}
 \begin{figure*}[ht]
\centering
\includegraphics[scale=.8]{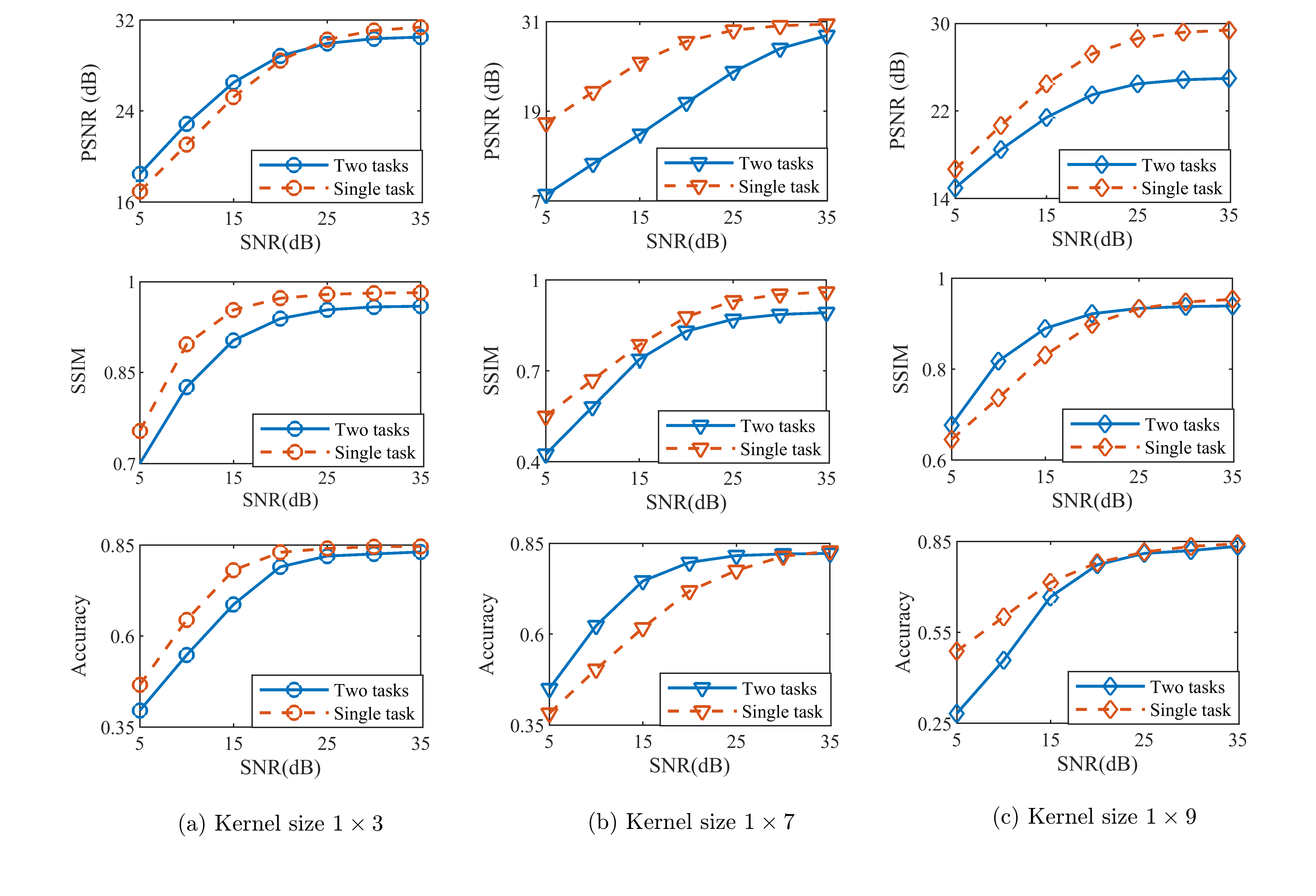}
\caption{Comparison of training a single task from scratch separately and fine-tuning two tasks in two stages under different kernel sizes. (The circle mark represents the performance with a kernel size of 1 $\times$ 3, the triangle mark represents 1 $\times$ 7, and the diamond mark represents 1 $\times$ 9.)}
\label{two-stage training}
\end{figure*}
\subsubsection{{ Performance Comparison under Different Scenarios}}

{  Table I offers a comparative analysis of performance across three distinct systems: the digital ODE-inspired network, the deep joint source-channel coding (DeepJSCC) framework, and the proposed analog Air-ODE network. In this setup, we integrate the Air-ODE block within the encoder block at the transmitter, thereby configuring the overall semantic communication system as the DeepJSCC framework \cite{bourtsoulatze2019deep}. The ODE-inspired network serves as the benchmark, achieving the highest PSNR and SSIM scores. Under 30 dB SNR conditions, DeepJSCC performance slightly decreases due to noise interference but remains robust. The proposed analog Air-ODE network experiences a slight performance drop attributed to noise interference and limited phase configurations, yet retains about 97\% of the performance of both the digital ODE network and DeepJSCC, demonstrating its resilience and efficiency under varying channel conditions. 

The Air-ODE block in the analog Air-ODE network operates at the speed of light during signal transmission, enabling computations to be completed in nanoseconds. In contrast, in the DeepJSCC and digital ODE-inspired neural networks, the ODE block computations rely on the baseband signal processor, requiring microseconds to milliseconds \cite{lin2018all}. In particular, in DeepJSCC, transmission delays still occur between the transceivers. However, our proposed analog Air-ODE network conducts computations directly during transmission, maintaining comparable performance while significantly reducing latency and enhancing computational efficiency.}

\subsubsection{SNR versus Performance}

\begin{table}[!h]
\caption{{\textbf{COMPARISON ON THE COMPLEXITY \\ UNDER DIFFERENT TRAINING METHODS}}}
\centering
\scalebox{0.75}
{
\begin{tabular}{c c c c c}
\hline
\specialrule{0em}{1pt}{1pt}
\hline
 \multicolumn{2}{c}{{{Training Methods}}} &   {{Training Cost}}  &  {{Inference Time}}  &  {{FLOPs}} \\
\midrule
 \multicolumn{2}{c}{{{Dual Tasks}}} &  {$N_{e}$  $\Delta_{\rm{DR}}$ + $N_{e_1}$   $\Delta_{\rm{EO}}$ +   $N_{e_2}$  $\Delta_{\rm{DS}}$} &   {5.06 ms}  &   {0.66 M}              \\
\midrule
\multirow{2}{*}{ {{Single Task}}} &{{Task\_{RI}}} &   {$N_{e}$($\Delta_{\rm{EO}}$ + $\Delta_{\rm{DR}}$)} &   {4.81  ms} &  {0.62 M}   \\
\cmidrule(l){2-5}
&{{Task\_{ST}}}  &   {$N_{e}$($\Delta_{\rm{EO}}$ + $\Delta_{\rm{DS}}$)}  &  {4.51 ms}   &  {0.36 M}  \\
\midrule
\specialrule{0em}{1pt}{1pt}
\hline
\end{tabular}}
\end{table}

Fig. \ref{SNR} illustrates the effects of varying SNR levels on the performance, with the digital ODE network serving as a benchmark unaffected by SNR variations. It explores the performance of the Air-ODE under different conditions, including scenarios with random phases and scenarios where the Air-ODE structure is absent.
It is worth noting that the performance of both image reconstruction and semantic tagging is significantly lower at all SNR levels when the Air-ODE framework is absent. This indicates that the lack of an Air-ODE framework has an obvious impact on the change of PSNR, SSIM and tagging accuracy. Though the performance increases slightly with the rise in SNR but remains significantly below ideal performance levels.

In the random phase scheme without proper configuration of the Air-ODE structure, moderate performance gains. This suggests that even a biased phase configuration can benefit from the inherent advantages of the ODE architecture. Additionally, the performance of our proposed analog Air-ODE improves with increasing SNR, approaching that of digital ODE networks at high SNR levels. These experimental results confirm the effectiveness of the analog Air-ODE in improving image reconstruction and semantic tagging quality under high SNR conditions.

\begin{figure*}[ht]
\centering
\includegraphics[scale=.65]{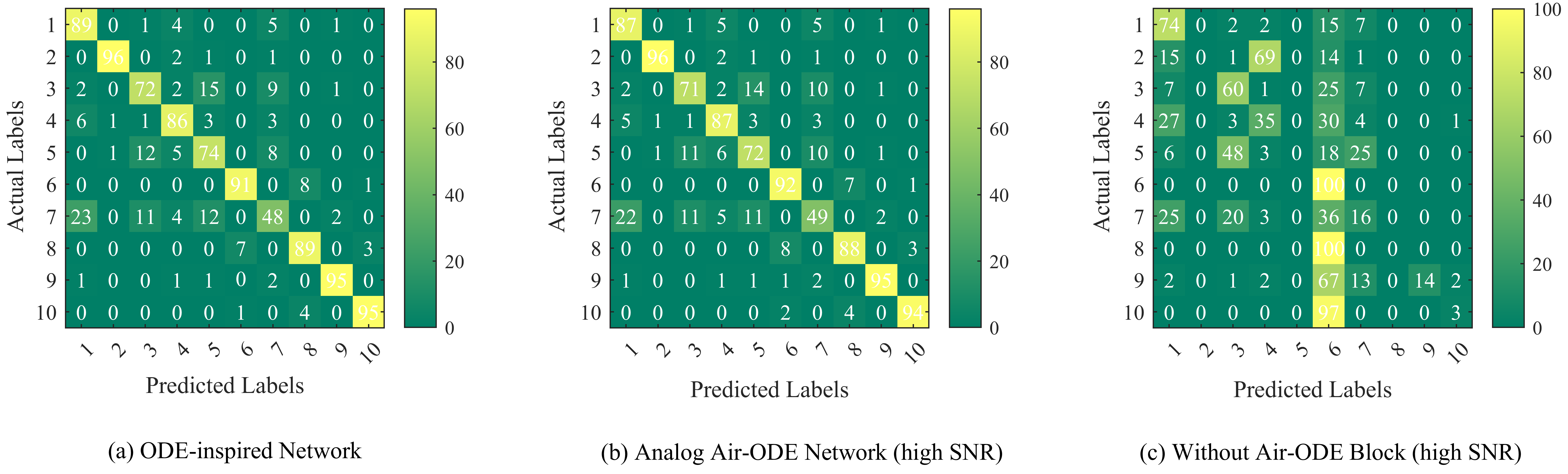}
\caption{Confusion matrix illustration under different networks. }
\label{Semantic tagging}
\end{figure*}

\subsubsection{Compress Ratio versus Performance}

\begin{figure}[ht]
\setlength{\abovecaptionskip}{0pt}
\setlength{\belowcaptionskip}{0pt}
\centering
\includegraphics[width= 0.45\textwidth]{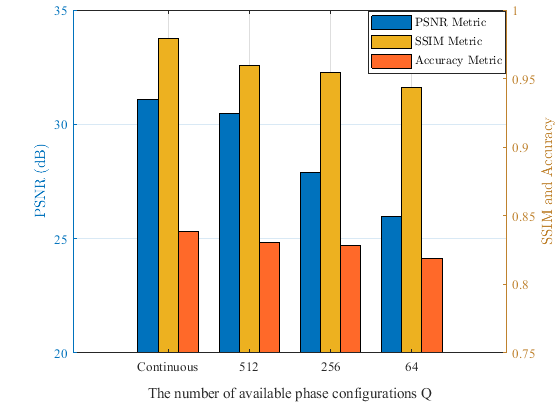}
\DeclareGraphicsExtensions.
\caption{{Performance versus the size of the feasible weight set.}}
\label{Quantization}
\end{figure}

Fig. \ref{compress} provides an analysis of the impact of various compression ratios on the PSNR, the SSIM, and the accuracy for both analog and digital ODE-inspired networks. The compression ratios examined include 64:1, 16:1, and 8:1. As illustrated in Fig. \ref{compress}, the performance of both the analog and digital ODE-inspired networks improves as the compression ratio decreases. This improvement is attributable to the increased amount of information that can be extracted and learned at lower compression ratios, thereby enhancing performance. In contrast, the setup without the analog Air-ODE structure performs the worst. The absence of a structured approach results in unstable performance, particularly under conditions of low compression, leading to unsatisfactory outcomes.

\subsubsection{Kernel Size versus Performance under Different Training Methods }
{  Table II and Fig. \ref{two-stage training} compare the network complexity and performance of the fine-tuning algorithm proposed in this paper against the conventional training-from-scratch approach.
In the fine-tuning method, the network first undergoes an initial training phase for \(N_{e_1}\) epochs, followed by a fine-tuning stage applied to a subset of the network for an additional \(N_{e_2}\) epochs, resulting in a total training duration of \(N_{e} = N_{e_1} + N_{e_2}\). 
In contrast, the training-from-scratch approach involves training two tasks independently, each for a total of \(N_{e}\) epochs, without leveraging prior knowledge from an initial phase.

Specifically, Table II compares the training costs, inference time, and floating point operations (FLOPs) of the two methods. In the fine-tuning approach, the first $N_{e_1}$ epochs involve the encoder-ODE cost $\Delta_{\rm{EO}}$ and the decoder cost for image reconstruction $\Delta_{\rm{DR}}$, denoted as $\Delta_{\rm{EO}} + \Delta_{\rm{DR}}$. In the second stage of $N_{e_2}$ epochs, training includes two decoder blocks, with a cost of $\Delta_{\rm{DR}} + \Delta_{\rm{DS}}$. 
For single-task training from scratch, each task is trained independently for $N_{e}$ epochs. As shown in Table II, the fine-tuning method reduces the training cost by $(N_{e} + N_{e_2}) \Delta_{\rm{EO}} + N_{e_1} \Delta_{\rm{DS}}$. Similarly, both the inference time and FLOPs in the fine-tuning method are lower than those in single-task training from scratch, with reductions of approximately 45.71\% and 32.65\%, respectively.
 }
 
 Training from scratch does not require balancing between two tasks, allowing it to better focus on the target task, thus training from scratch outperforms fine-tuning in Fig. \ref{two-stage training}. However, the performance difference between the two methods is minimal, demonstrating the effectiveness of our fine-tuning algorithm. In all the results, the performance of networks deployed under both training methods improves with increasing SNR. This is because higher SNR results in less noise interference, leading to the performance being closer to that of the digital ODE-inspired neural network.
Additionally, when considering different kernel sizes in the Air-ODE blocks, it is observed that smaller kernels, such as 1 $\times$ 3, provide better performance compared to larger kernels. This is because smaller convolution kernels can more finely capture local features in the image,  which are crucial for image reconstruction and tagging tasks. Smaller kernels are better at extracting these details, enhancing the overall performance of the network.

\subsubsection{Confusion Matrix for the Semantic Tagging Task}

The results for semantic tagging under three different network configurations are shown in Fig. \ref{Semantic tagging}, where the data in the matrices represent the percentage of actual data.
In Fig. \ref{Semantic tagging}(a), the performance of the ODE-inspired neural network is high across most tagging, except for label 1 and label 7, due to their similar shape\footnote{Although the overall accuracy of the ODE-inspired network may be improved by optimizing the model, this paper only focuses on the differences between digital and deployed analog networks rather than on optimizing the digital network's performance.}.
Fig. \ref{Semantic tagging}(b) illustrates the analog Air-ODE network under high SNR conditions, the tagging results closely match those of the digital ODE-inspired networks. Interestingly, there is a slight improvement in the correct classification of categories 6 and 7, which can be attributed to the beneficial effects of small amounts of noise. Fig. \ref{Semantic tagging}(c) shows the tagging results for the analog networks without the Air-ODE block. Despite the high SNR, the results are significantly affected, with a noticeable reduction in the diagonal elements, which underscores the necessity of incorporating the Air-ODE block.
By analyzing the confusion matrices, we can verify the specific influence of channel noise on the semantic tagging task and emphasize the critical role of Air-ODE blocks in maintaining tagging accuracy.

\subsubsection{The Size of Feasible Weight Set versus Performance}

Then, Fig. \ref{Quantization} depicts the performance variations of the analog Air-ODE network under different quantization conditions, specifically comparing ideal continuous phases (where the number of feasible weights is virtually infinite) to quantized phases with $Q$ values of 512, 256, and 64. The results clearly show that both the PSNR and the SSIM experience degradation as the quantization set becomes more restricted. The comparison between continuous and quantized phases illustrates that only less than 10$\%$ loss is observed when practical discrete phase shift constraint is considered, showing the robustness of the proposed scheme.

\section{Conclusion}

In this paper, we introduce a distributed RIS-aided communication system designed to perform image reconstruction and semantic tagging tasks simultaneously. The proposed approach deploys complex-valued Air-ODE networks within distributed RIS-aided communication environments. The utilization of distributed RISs not only enhances spectrum efficiency but also reduces energy consumption by offloading computational tasks to the environment. Simulation results demonstrate that the analog Air-ODE network can achieve satisfactory performance in recovering high-quality images and predicting the corresponding tagging results from the compressed features. Significant gains are observed in PSNR, SSIM, and tagging accuracy metrics, especially at high SNR levels, due to the Air-ODE structure. These findings highlight the potential for future advancements in integrating wireless communication environments with AI technologies. In future work, we plan to explore a broader range of network frameworks and incorporate more RISs to enhance our system's capabilities for handling increasingly complex tasks and larger datasets.

\bibliographystyle{IEEEtran}
\bibliography{IEEEabrv,myref}

\end{document}